\documentclass[journal=jctcce,manuscript=article,layout=traditional]{achemso} 

\usepackage[T1]{fontenc} 
\usepackage[version=3]{mhchem} 
\usepackage[dvipsnames]{xcolor}
\usepackage{graphicx,subcaption}
\usepackage{amssymb}
\usepackage{overpic}
\usepackage{soul}
\usepackage{bm}
\usepackage{ragged2e}
\usepackage{array, multirow, boldline}

\newif\ifhighlightchanges
\highlightchangesfalse

\definecolor{darkblue}{HTML}{003D6D}

\newcommand{\bP}{{\bm P}}
\newcommand{\bF}{{\bm F}}
\newcommand{\br}{{\bm r}}
\newcommand{\bp}{{\bm p}}

\newcommand{\bd}{{\bm d}}
\newcommand{\bM}{{\bm M}}

\newcommand{\bGamma}{{\bm \Gamma}}

\newcommand{\hH}{\hat H}
\newcommand{\hT}{\hat T}
\newcommand{\hE}{\hat E}
\newcommand{\hp}{\hat p}
\newcommand{\hL}{\hat L}

\newcommand{\braket}[3]{\left< #1\left|#2\right|#3 \right>}

\newcommand{\bra}[1]{\left< #1\right|}
\newcommand{\ket}[1]{\left|#1 \right>}

\usepackage{accents}

\author{Zhen Tao}
\affiliation{Department of Chemistry, University of Pennsylvania, Philadelphia, Pennsylvania 19104, USA.}
\author{Tian Qiu}
\affiliation{Department of Chemistry, University of Pennsylvania, Philadelphia, Pennsylvania 19104, USA.}
\author{Mansi Bhati}
\affiliation{Department of Chemistry, University of Pennsylvania, Philadelphia, Pennsylvania 19104, USA.}
\author{Xuezhi Bian}
\affiliation{Department of Chemistry, University of Pennsylvania, Philadelphia, Pennsylvania 19104, USA.}
\author{Titouan Duston}
\affiliation{Department of Chemistry, University of Pennsylvania, Philadelphia, Pennsylvania 19104, USA.}
\author{Jonathan Rawlinson}
\affiliation{Department of Mathematics, University of Manchester, Manchester M13 9PL, UK}
\author{Robert G. Littlejohn}
\affiliation{Department of Physics, University of California, Berkeley, California 94720, USA}
\author{Joseph E. Subotnik}
\email{subotnik@sas.upenn.edu}
\affiliation{Department of Chemistry, University of Pennsylvania, Philadelphia, Pennsylvania 19104, USA.}

\title[]{Practical Phase-Space Electronic 
Hamiltonians for {\em Ab Initio} Dynamics}

\begin{document}

\begin{abstract}
    Modern electronic structure theory is built around the Born-Oppenheimer approximation and the construction of an electronic Hamiltonian $\hH_{el}(\bm X)$ that depends on the nuclear position $\bm X$ (and not the nuclear momentum $\bP).$  In this article, using the well-known theory of electron translation ($\bGamma'$) and rotational ($\bGamma''$) factors to couple electronic transitions to nuclear motion,  we construct a practical phase-space electronic Hamiltonian that depends on both nuclear position and momentum, $\hH_{PS}(\bm X,\bP)$. While classical Born-Oppenheimer dynamics that runs along the eigensurfaces of the operator  $\hH_{el}(\bm X)$  can recover many nuclear properties correctly, we present some evidence that motion along the eigensurfaces of $\hH_{PS}(\bm X,\bP)$ can better capture both nuclear {\em and electronic} properties (including the elusive electronic momentum studied by Nafie). Moreover, only the latter (as opposed to the former) conserve the {\em total} linear and angular momentum in general.
\end{abstract}

\section{Introduction}
Born-Oppenheimer (BO) theory\cite{Born1927} is the work horse of molecular dynamics\cite{Tully2000,Barbatti2011,Curchod2013,Furche2013:fsshreview,tretiak:2014:acr}. The basic premise is that one separates the slow nuclear degrees of freedom ($\bm X$) from the fast electronic degrees of freedom ($\bm r$).  Mathematically, one decomposes the total  Hamiltonian (nuclear + electronic, $\hH_{ tot}$)  into the nuclear kinetic energy operator $\hT_{ n}(\bP)$ and the non-relativistic electronic Hamiltonian $\hH_{ e}(\bm X)$
\begin{align}
    \hH_{tot} = \hT_{ n}(\bP) + \hH_{ e}(\bm X)\label{eq:H_BO}
\end{align}
For the moment, we ignore  external fields.
If one is running classical BO dynamics, one then diagonalizes the resulting electronic Hamiltonian,
\begin{align}
    \hH_{e}(\bm X) \ket{\Phi_I(\bm X)}=E_I(\bm X)\ket{\Phi_I(\bm X)}\label{eq:psi_BO}
\end{align}
and propagates Newton's equations along the  eigensurface $E_I(\bm X)$:
\begin{align}
    \dot{\bm X}&=\frac{\bP}{\bM} \\
     \dot{\bP}&=-\frac{\partial E_I}{\partial \bm X}   \label{eq:dP_BO}
\end{align}
Here, our notation is as follows. We label all electronic operators (or more generally matrices) with a hat, {\bf boldface} indicated vectors in three dimensional space, and we index all standard adiabatic states by $I, J, K, \ldots$.  Atomic orbitals (AO) are indexed below by $\{\mu,\nu,\lambda,\sigma\}$. The nuclei are indexed below by $A,B,C,Q$ and the Cartesian $xyz$ are indexed by $\alpha\beta\gamma\kappa\eta..$

The Hamiltonian dynamics in Eqs. \ref{eq:H_BO} - \ref{eq:dP_BO} conserve the total energy, the nuclear linear momentum, and the nuclear angular momentum. However, as demonstrated recently, these calculations do not conserve the {\em total} linear or angular momentum in general\cite{Bian2023}.  One can easily reach this conclusion by imagining that we run BO dynamics along one doublet surface here as constructed for a system with an odd number of electrons and spin-orbit coupling (SOC).  In such a case,the electronic wavefunction becomes complex-valued  and it is clear that the electronic linear and angular momentum of a given state will be nonzero, changing as a function of time. However, Eqs. \ref{eq:H_BO} - \ref{eq:dP_BO} will conserve only the nuclear  angular momentum; thus, the {\em total} angular momentum must change as a function of time.

Given that the total angular momentum operator is strictly diagonal in the BO representation and should formally be conserved if one runs quantum dynamics along a single BO surface with the correct phase conventions \cite{Littlejohn2023}, the resolution of this paradox is that 
one must supplement Eq. \ref{eq:dP_BO} with the proper Berry force:\cite{Bian2023}
\begin{eqnarray}
\label{eq:pi_dot}
     \dot{\bP}&=&-\frac{\partial E_I}{\partial \bm X}    + \bF^{\text{Berry}}_{I}
     \\
 \label{eqn:berry_force}
    \bF^{\text{Berry}}_{I} &=& \bm \Omega_{I} \cdot \dot{\bm{ X}} \\
    \label{eqn:ab_BC}
    \bm \Omega_{I} & = & i\hbar\nabla_{n} \times \bm{d}_{II}
\end{eqnarray}
Here $\bm \Omega$ is the abelian Berry curvature and $\bd_{II} = \bra{\Phi_{I}}\nabla_{n}\ket{\Phi_{I}}$ is the on-diagonal derivative coupling vector between the state $I$ and itself. The quantity $\bm A = i\hbar\bm d_{II}$ is also known as the Berry connection and it is non-zero for systems with odd number of electrons\cite{Mead1979} and SOC for which we have complex wavefunctions and Kramers' pairs.  By including the Berry force in Eq. \ref{eq:pi_dot}, one recovers total angular momentum conservation with classical nuclear dynamics\cite{Bian2023}.

Unfortunately, however, including a Berry force has two pitfalls. $(i)$ Computing a Berry force can be quite expensive numerically, because the Berry curvature involves the derivative of a derivative coupling (i.e. a second derivative)--which must always be avoided in {\em ab initio} calculations whenever possible. Note that the calculation can be made simpler for the specific case of a ground state calculation where the tensor has been derived and implemented for a  generalized Hartree-Fock (HF) ansatz,\cite{Culpitt2022,Tao2023} but overall it is certainly desirable to avoid the Berry curvature for excited state dynamics. $(ii)$ For the case of a system with electronic degeneracy (i.e.  a Kramer's doublet), the notion of a Berry force is tricky and somewhat arbitrary because one must arbitrarily pick one doublet (e.g. $I$) out of a pair (e.g. $I,I'$). One can avoid the second pitfall by running Ehrenfest (mean-field) dynamics,\cite{Ehrenfest2023} calculating the electronic density matrix $\hat{\rho}$ over a set of states and, as prescribed  by Takatsuka \cite{Amano2005} and then Krishna \cite{Krishna2007}, including the non-Abelian curvature\cite{mead:1992:rmp} in the force:
\begin{eqnarray}
\label{eqn:non_ab_BC}
    \hat{\Omega}^{B\alpha C\beta}_{\text{nab}} =    \frac{\partial  {{\hat {A}}^{C\beta}}}{\partial X_{B\alpha}} - 
     \frac{\partial  {{\hat {A}}^{B\alpha}}}{\partial X_{C\beta}} -
     \frac{i}{\hbar} \left[ {{\hat {A}}^{B \alpha}}, {{\hat {A}}^{ C \beta}}
    \right]
\end{eqnarray}
In such a case, one never needs to pick an arbitrary surface within a Kramers' pair and one does recover total (linear and angular) momentum conservation\cite{Ehrenfest2023}. Alas, however, computing the non-Abelian Berry curvature in Eq. \ref{eqn:non_ab_BC} is even more expensive than computing the Abelian Berry curvature in Eq. \ref{eqn:ab_BC}, and so the method will likely need future approximations to be practical for large systems.

\subsection{Phase-Space Electronic Hamiltonians}
Now, apart from including a Berry force of any kind, an alternative  means to  conserve the total momentum is to work from the start with a phase-space electronic Hamiltonian that depends on both nuclear position and momentum, $\hH_{PS}(\bm X,\bP)$:
\begin{align}
  \label{HPS}
   \hH_{\rm PS}(\bm X,\bP) &= \frac{\bP^2}{2\bM} - i\hbar \frac{\bP}{\bM} \cdot \hat{\bm \Gamma} + \hH_{\rm el}(\bm X) ,
\end{align}
Here, we emphasize that the nuclear momentum operator has been replaced by the classical momentum $\bm P$ and $\hat{\bm \Gamma}$ is a nuclear-electronic coupling to be determined.\cite{Wu2023} 
The most famous example of Eq. \ref{HPS} is Shenvi's phase-space electronic Hamiltonian\cite{Shenvi2009} in an adiabatic electronic basis:
\begin{align}
  \label{Hshenvi}
   \hH_{\rm Shenvi}(\bm X,\bP) &= \frac{\bP^2}{2\bM}   - i\hbar \frac{\bP \cdot \hat{\bm d}}{\bM} - \hbar ^2 \frac{\hat{\bm d} \cdot \hat{\bm d} }{2 \bM} + \hE_{ad}(\bm X)
\end{align}
Here, $\hE_{ad}$ is the diagonal matrix of eigenvalues of $\hH_{el}(\bm X)$, and $\hat{\bm d}$ is the density matrix of derivative couplings that couples different electronic states $I,J$, where
\begin{align}
    \bP\cdot\hat{\bd} &= \sum_{A\alpha}P_{A\alpha}\hat{d}^{A\alpha}\\
    \hat{\bm d}\cdot\hat{\bm d}&= \sum_{A\alpha}\hat{d}^{A\alpha}\hat{d}^{A\alpha}\\
    d_{IJ}^{A \alpha} &= \left< \Phi_I \middle|  \frac{\partial
}{\partial X_{A \alpha}} \Phi_J \right>.\label{eq:deriv_Mat}
\end{align}
 As shown in Ref. \citenum{Wu2023}, if we diagonalize  $\hH_{Shenvi}(\bm X,\bP)$ and run classical dynamics along the corresponding eigensurfaces (that depend on $\bm X$ and $\bm P$), we will indeed conserve the total linear and angular momentum.

Alas, there are also problems with the Hamiltonian in Eq. \ref{Hshenvi}, some conceptual and some practical.
$(i)$ First, one limitation arises in the case of degenerate states, e.g. doublets. In such a case, given the complex-valued nature of the SOC, one must arbitrarily choose two basis states  ($\ket{\Phi_I}$ and $\ket{\Phi_{I'}}$) and two complex-valued phase for each state, which leads to an arbitrary $\hat{\bm d}$ in Eq. \ref{Hshenvi}, which ultimately renders the phase-space Hamiltonian of limited value in such a case; the algorithm cannot be reliable for systems with electronic degeneracy, e.g. systems with odd numbers of electrons.
$(ii)$ A second limitation  is numerical stability. Near a conical intersection, the derivative coupling $\hat{\bm d}$ will diverge, and one will recover spikes in the eigenenergies of $\hH_{\rm Shenvi}(\bm X,\bP)$ in Eq. \ref{Hshenvi}.\cite{izmaylov:2016:jpc_dboc_pssh} 
$(iii)$ A third limitation is computational cost.  The algorithm requires a complete set of derivative couplings to construct $\hH_{\rm Shenvi}(\bm X,\bP)$ in Eq. \ref{Hshenvi}, and as discussed above,  running dynamics on the potential energy surfaces requires the derivatives of those derivative couplings.
For these reasons, as far as we are aware, no one has ever worked with the Hamiltonian in Eq. \ref{Hshenvi} for any {\em ab initio} calculations. Thus it is desirable to approximate the derivative coupling vector while maintaining important features such as momentum conservation.

\subsection{Necessary Conditions for Linear and Angular Momentum Conservation}
Interestingly, to satisfy momentum conservation, Eq. \ref{Hshenvi} is not the only phase-space Hamiltonian. In fact, if all ones seeks is momentum conservation, one does not need to compute $\hat{\bm d}$ necessarily. As discussed in Ref. \citenum{Wu2023} in the phase-space surface hopping context, if one works with a subspace of adiabatic states and introduces a $\Gamma^{A\alpha}_{IJ}$ matrix (in place of the derivative coupling), one can ensure momentum conservation as long as the following four conditions are satisfied:
\begin{eqnarray}
    -i\hbar \sum_{A}\Gamma^{A\alpha}_{IJ} + \bra{\Phi_{I}}\hat{p}^{\alpha}_{e}\ket{\Phi_{J}} & =  0 \label{eq:condition1}\\
    \sum_{B} \nabla_{B\beta}\Gamma^{A\alpha}_{IJ} & =  0 \label{eq:condition2}\\    
    -i\hbar \sum_{A\beta\gamma} \epsilon_{\alpha\beta\gamma} X_{A\beta} \Gamma^{A\gamma}_{IJ} + \bra{\Phi_{I}}\hat{L}^{\alpha}_{e} + \hat{s}^{\alpha}\ket{\Phi_{J}} &= 0 \label{eq:condition3}\\
    \sum_{B\beta\eta}\epsilon_{\alpha\beta\eta} X_{B\beta} \nabla_{B\eta}\Gamma^{A\gamma}_{IJ} + \sum_{\eta}\epsilon_{\alpha\gamma\eta} \Gamma^{A\eta}_{IJ} &= 0\label{eq:condition4}
\end{eqnarray}
Here $\epsilon_{\alpha\beta\gamma}$ is the Levi-Civita symbol. Below, we will write the matrix elements of $\hat{p}^{\alpha}_{e}$ and $\hat{L}^{\alpha}_{e}$ in AO basis as $p^{\alpha}_{\mu\nu}$ and $l^{\alpha}_{\mu\nu}$, respectively.

On the one hand, the conditions in Eq. \ref{eq:condition1} and Eq. \ref{eq:condition3} are really phase conventions,\cite{Littlejohn2023} i.e. the conventions for choosing the phases of the electronic states at different geometries.  On the other hand, the conditions in Eq. \ref{eq:condition2} and Eq. \ref{eq:condition4} ensure that the the couplings transform are invariant under translational and rotational changes of coordinate.  
\begin{align}
    \Gamma^{A \alpha}_{IJ}(\bm X_0 + { \delta \bm X}) &= \Gamma^{A \alpha}_{IJ}(\bm X_0) 
    \tag{\ref{eq:condition2}$'$} \label{eq:condition2prime}
    \\
    \bGamma_{IJ}(\bm{R} \bm X_0) &= \bm{R}\bGamma_{IJ}(\bm X_0)     \tag{\ref{eq:condition4}$'$} \label{eq:condition4prime}
\end{align}
Here $\bm{R}$ is the rotation operator in the Cartesian $xyz$ space, $\bm{R} = \mathrm{exp}(-\frac{i}{\hbar}\sum_{\alpha}\bm{L}^{\alpha}\delta_{\alpha})$ and $L^{\alpha}_{\beta\gamma} = i\hbar \epsilon_{\alpha\beta\gamma}$. 
As a side note, as shown in Ref. \citenum{Wu2023},  the full derivative couplings $\hat{\bm d}$ do satisfy all four constraints (Eqs. \ref{eq:condition1}-\ref{eq:condition4}). 

Now, with so many choices for $\bGamma_{IJ}$ possible in order to conserve momentum, one must wonder what is the optimal path for building such a term. For practical purposes, one would prefer the simplest way possible. In particular, one would like to avoid the cumbersome process of diagonalizing the electronic Born-Oppenheimer Hamiltonian, generating states $I$ and $J$, adding  a momentum-dependent term ${\bm P} \cdot \bGamma_{IJ}$ in the spirit of Eqs. \ref{eq:H_BO}-\ref{eq:dP_BO},  rediagonalizing the resulting phase-space Hamiltonian, and then generating gradients for dynamics.
Such a multi-step approach is not very 
practical.

\subsection{Outline of this paper}
The goal of this paper is to provide an alternative, one-shot treatment for generating phase-space Hamiltonians with $\bGamma$-couplings. More precisely, we will derive and implement a meaningful $\bm \Gamma$ matrix that arises from a one-electron operator so that, at the end of the day, we need only to diagonalize a single electronic Hamiltonian, and motion along the resulting surfaces will automatically conserve momentum while also being very efficient. 

To that end, in Sec. \ref{sec:theory} below, we will show that Eqs. \ref{eq:condition1}-\ref{eq:condition4} for $\bGamma_{IJ}$ can indeed be satisfied  for a proper one-electron operator $\bm\Gamma_{\mu \nu} a^{\dagger}_{\mu} a_{\nu}$ in an atomic orbital basis, $ \bGamma_{IJ} = 
\sum_{\mu \nu} \left< \Phi_I \middle| \bm\Gamma_{\mu \nu} a^{\dagger}_{\mu} a_{\nu} \middle| \Phi_J \right>$, and we will delineate the necessary conditions for $\bm\Gamma_{\mu \nu}$  in order to conserve linear and  angular momentum.  Next, with so many possible choices for $\bGamma$, in Sec. \ref{sec:nafie}, we will posit that one means for isolating a physically meaningful  one-electron operator $\bm\Gamma_{\mu \nu}$ is to insist that the phase-space electronic Hamiltonian should yield the correct expression for linear  (and sometimes angular) momentum (which requires a beyond Born-Oppenheimer calculations as developed by Nafie\cite{Nafie1983,Nafie1992,Nafie1997,Nafie2004}). 
 Finally, in Sec. \ref{sec:result}, we provide results, demonstrating both the size of the relevant matrix elements as well as the capacity of the resulting method to recapitulate electronic momentum and angular momentum. In Sec. \ref{sec:discussion}, we discuss the future implications of this work and conclude.

\section{Theory: A Stable Ansatz For a One-Electron Hamiltonian Operator $\Gamma_{\mu \nu}$}\label{sec:theory}
While Eqs. \ref{eq:condition1}-\ref{eq:condition4} dictate the form of $\bm \Gamma_{IJ}$ in a many-body basis, the relevant conditions required of a single electron operator $\bGamma_{\mu \nu}$ are slightly different because the orientation of each atomic orbitals is always fixed in the lab frame and does not rotate with the molecule. In other words, for example, for a rotation molecule, a $p_x$ orbital is always polarized in the $x-$direction.  Nevertheless, we will show below that if the following conditions are obeyed, 
 \begin{align}
 \label{eq:Gamma_uv1}
     -i\hbar\sum_{A}\Gamma^{A\alpha}_{\mu\nu} + p^{\alpha}_{\mu\nu} &= 0  \\
      \label{eq:Gamma_uv2}
\sum_{B}\nabla_{B\beta}\Gamma^{A\alpha}_{\mu\nu} &= 0\\
     \label{eq:Gamma_uv3}
    -i\hbar\sum_{A\beta\gamma}\epsilon_{\alpha\beta\gamma}{X}_{A\beta} \Gamma^{A\gamma}_{\mu\nu} + l^{\alpha}_{\mu\nu} + s^{\alpha}_{\mu\nu} &= 0
    \\
     \sum_{B\beta\eta} \epsilon_{\alpha\beta\eta} X_{B\beta} \bra{\mu} \frac{\partial \hat{\Gamma}^{A\gamma}}{\partial X_{B\eta}}\ket{\nu} -\frac{i}{\hbar}
        \bra{\mu} [\hat{\Gamma}^{A\gamma},\hat{L}^{\alpha}_{e} ]\ket{\nu} + \sum_{\eta}\epsilon_{\alpha\gamma\eta} \Gamma^{A\eta}_{\mu\nu} &= 0 \label{eq:Gamma_uv4}
 \end{align} 
 then the linear and angular momentum will be conserved. Similar to the many-body case above, we note that 
 Eqs. \ref{eq:Gamma_uv2} and \ref{eq:Gamma_uv4} are equivalent to 
  \begin{align}
      \tag{\ref{eq:Gamma_uv2}$'$}\label{eq:Gamma_uv2prime}
    \Gamma^{A \alpha}_{\mu \nu}(\bm X_0 + { \delta \bm X}) &= \Gamma^{A \alpha}_{\mu \nu}(\bm X_0) \\
    \tag{\ref{eq:Gamma_uv4}$'$}
     \label{eq:Gamma_uv4prime}
    \bGamma_{\bar{\mu} \bar{\nu}}(\bm{R} \bm X_0) &= \bm{R}\bGamma_{\mu \nu}(\bm X_0)
 \end{align} 
Here, a word about notation and basis set definition is essential. If $\ket{\nu}$ is centered on atom $B$, and the molecule is rotated by an amount $\bm{\delta},$ then here we define the atomic orbital $\ket{\bar{\nu}} $ in Eq. \ref{eq:Gamma_uv4prime} to be that orbital generated by rotation around atom $B$ by the same rotational angle, $\ket{\bar{\nu}_{B}} \equiv \mathrm{exp}(-\frac{i}{\hbar}\sum_{\alpha}\hat{L}^{B\alpha}_{e}\delta_{\alpha})\ket{\nu_{B}}$. 
As mentioned above, the fact that atomic orbitals are defined in a lab frame (rather than molecular frame) within quantum chemistry codes leads to the differences between Eqs. \ref{eq:Gamma_uv4} and \ref{eq:Gamma_uv4prime} with Eqs. \ref{eq:condition4} and \ref{eq:condition4prime} above (where  the indices of $\bGamma(\bm{R} \bm X_0)$
are $\bar{\mu},\bar{\nu}$ instead of $\mu,\nu$). The equivalence between  Eq. \ref{eq:Gamma_uv4prime} and  Eq. \ref{eq:Gamma_uv4} for a general one-electron operator is shown in  Appendix \ref{app:ao_rot}.

\subsection{Equations of Motion}
Let us now prove the claims above vis a vis momentum conservation. We
begin by writing the total energy of the phase-space Hamiltonian for a single state in an  atomic orbital basis:\begin{eqnarray}
\label{eq:pssh_e}
        E_{\rm PS}(\bm X,\bP) &=&\frac{\bP^2}{2\bM}+ \sum_{\mu\nu}D_{\nu\mu} \Big(h_{\mu\nu} - i\hbar \sum_{A\alpha} \frac{ P_{A\alpha}\Gamma^{A\alpha}_{\mu\nu}}{M^{A}}\Big) + \sum_{\mu\nu\lambda\sigma}G_{\nu\mu\sigma\lambda}\pi_{\mu\nu\lambda\sigma} 
\end{eqnarray}
Here the quantities $D_{\nu\mu}$ and $G_{\nu\mu\sigma\lambda}$ are the one-electron and two-electron density matrix elements obtained from solving the phase-space Hamiltonian directly. The $h_{\mu\nu}$ and  $\pi_{\mu\nu\lambda\sigma} $ matrix elements are the relevant one-electron and two-electron operators in the atomic orbital basis.  Given this energy expression, we can write the classical equations of motion to propagate the phase-space variables $(\bm X,\bP)$.
\begin{eqnarray}
    \dot{X}_{A\alpha}&=&\frac{P_{A\alpha}}{M^{A}} - i\hbar\sum_{\mu\nu}\frac{D_{\nu\mu}\Gamma^{A\alpha}_{\mu\nu}}{M^{A}} \label{eq:dR_ps}\\
     \dot{P}_{A\alpha}&=&-\frac{\partial E_{PS}}{\partial X_{A\alpha}}  \label{eq:dP_ps1} \\
     &=&-\sum_{\mu\nu}D_{\nu\mu}\Big(\frac{\partial h_{\mu\nu}}{\partial X_{A\alpha}} -\sum_{B\beta}\frac{i\hbar P_{B\beta}}{M^{B}} \frac{\partial \Gamma^{B\beta}_{\mu\nu} }{\partial X_{A\alpha}}  \Big) - \sum_{\mu\nu\lambda\sigma}G_{\nu\mu\sigma\lambda}\frac{\partial \pi_{\mu\nu\lambda\sigma} }{\partial X_{A\alpha}} + \sum_{\mu\nu}W_{\nu\mu}\frac{\partial S_{\mu\nu}}{\partial X_{A\alpha}}\label{eq:dP_ps2}
\end{eqnarray}
Here $\bm S$ matrix is the overlap matrix. In going from Eq. \ref{eq:dP_ps1} to Eq. \ref{eq:dP_ps2}, we avoid the expensive step of calculating the orbital dependence on the nuclear positions according to the Wigner's (2n+1) rule (which is well known in quantum chemistry).\cite{Wigner1935,Ángyán2009}  Now let us examine the momentum conservation laws. 

\subsection{Linear Momentum} 
For the case of linear momentum conservation, using Eq. \ref{eq:dR_ps}, one evaluates:
\begin{eqnarray}
    \frac{d}{dt} P^{\alpha}_{tot}& =&\frac{d}{dt} \Big( \sum_{A} M_{A} \dot{X}_{A\alpha} +  \sum_{\mu\nu} D_{\nu\mu}p^{\alpha}_{\mu\nu} \Big) \label{eq:dpdt_1}\\
    & = & \frac{d}{dt} \Bigg[ \sum_{A} P_{A\alpha} +  \sum_{\mu\nu} D_{\nu\mu}\Big( \sum_A -i\hbar\Gamma^{A\alpha}_{\mu\nu} + p^{\alpha}_{\mu\nu} \Big)\Bigg] \label{eq:dpdt_2}
\end{eqnarray}

Now, if we plug Eq. \ref{eq:Gamma_uv1} into Eq. \ref{eq:dpdt_2}, the second term on the right hand side (RHS) vanishes and we can simplify:
\begin{eqnarray}\label{eq:dpdt_3}
    \frac{d}{dt} P^{\alpha}_{tot} = \sum_{A} \dot{P}_{A\alpha}
    = -\sum_{A}\frac{\partial E_{PS}}{\partial X_{A\alpha}}
\end{eqnarray}
From Eq. \ref{eq:dpdt_3}, one concludes that in order for the total linear momentum to conserve ($\frac{d}{dt} P^{\alpha}_{tot} = 0$), the phase-space energy needs to be translationally invariant ($\sum_{A}\frac{\partial E_{PS}}{\partial X_{A\alpha}}  = 0$). If we dig a little deeper and plug Eq. \ref{eq:dP_ps2} into Eq. \ref{eq:dpdt_3}, we find
\begin{eqnarray}
    \frac{d}{dt} P^{\alpha}_{tot}   
    & = & i\hbar\sum_{B\beta\mu\nu} \frac{P_{B\beta}}{M^{B}} D_{\nu\mu}\sum_{A} \frac{\partial \Gamma^{B\beta}_{\mu\nu} }{\partial X_{A\alpha}} \label{eq:dpdt_4}
\end{eqnarray}
Note that, in Eq. \ref{eq:dpdt_4}, we have already excluded the relevant  $h_{\mu\nu}$, $ \pi_{\mu\nu\lambda\sigma}$, and $ S_{\mu\nu}$ terms from the gradient  because these matrices are translation-invariant, i.e. $\sum_{A}\frac{\partial h_{\mu\nu}}{\partial X_{A\alpha}}  = 0$, $\sum_{A}\frac{\partial \pi_{\mu\nu\lambda\sigma}}{\partial X_{A\alpha}}  = 0$, $\sum_{A}\frac{\partial S_{\mu\nu}}{\partial X_{A\alpha}}  = 0$. Finally, we conclude that according to Eqs. \ref{eq:Gamma_uv2} and \ref{eq:dpdt_4}, $\frac{d}{dt} P^{\alpha}_{tot}  = 0$. Thus, conditions Eqs. \ref{eq:Gamma_uv1}-\ref{eq:Gamma_uv2} are sufficient to guarantee linear momentum conservation.

\subsection{Angular Momentum} 
Next, we evaluate the angular momentum:
\begin{eqnarray}
    \frac{d}{dt} L^{\alpha}_{tot} & =& \frac{d}{dt} \Big[ \sum_{A\beta\gamma} \epsilon_{\alpha\beta\gamma}X_{A\beta}M_{A} \dot{X}_{A\gamma} + \sum_{\mu\nu} (l^{\alpha}_{\mu\nu} + s^{\alpha}_{\mu\nu})  D_{\nu\mu}  \Big]  \label{eq:dldt_1}\\
    & = &  \frac{d}{dt}  \Bigg[\sum_{A\beta\gamma}  \epsilon_{\alpha\beta\gamma}{X}_{A\beta}{P}_{A\gamma} + \sum_{A\mu\nu\beta\gamma}    \Big(-i\hbar\epsilon_{\alpha\beta\gamma}{X}_{A\beta}\Gamma^{A\gamma}_{\mu\nu} + l^{\alpha}_{\mu\nu} + s^{\alpha}_{\mu\nu}\Big) D_{\nu\mu}  \Bigg]  \label{eq:dldt_2}
\end{eqnarray}
If we plug Eq. \ref{eq:Gamma_uv3} into Eq. \ref{eq:dldt_2} to eliminate the second term on the RHS of Eq. \ref{eq:dldt_2}, we are left with
\begin{eqnarray}
    \frac{d}{dt} L^{\alpha}_{tot} & =& \sum_{A\beta\gamma} \epsilon_{\alpha\beta\gamma}\Big(\dot{X}_{A\beta}{P}_{A\gamma} + X_{A\beta}\dot{P}_{A\gamma} \Big)  \label{eq:dldt_3}\\
     & = & - \sum_{A\beta\gamma} \epsilon_{\alpha\beta\gamma}\Big({P}_{A\beta}\frac{\partial E_{PS}}{\partial P_{A\gamma}} + {X}_{A\beta}\frac{\partial E_{PS}}{\partial X_{A\gamma}}\Big) \label{eq:dldt_4}
\end{eqnarray}
From Eq. \ref{eq:dldt_4}, it follows (not surprisingly) that in order for the total angular momentum to be conserved, the energy of the phase-space electronic Hamiltonian must be rotationally invariant. Mathematically, requiring a vanishing RHS of Eq. \ref{eq:dldt_4} is equivalent to requiring the energy $E_{PS} (\bm X,\bP) $ in Eq. \ref{eq:pssh_e} to satisfy:
\begin{equation}
    E_{PS} (\bm{R}\bm X,\bm{R}\bP) = E_{PS} (\bm X,\bP) 
    \label{eq:epsrot}
\end{equation}

Let us now evaluate all of the terms in Eq. \ref{eq:pssh_e} individually. To begin with, one can immediately see the nuclear kinetic energy is rotationally invariant because $\bP^2 = \bm{R}\bP\cdot\bm{R}\bP$. For the remaining contributions in  Eq. \ref{eq:pssh_e}, we recall that  the one-electron and the two electron operators written in the atomic orbital basis satisfy  

\begin{eqnarray}
\label{eq:1e_rot}
    h_{\bar{\mu}\bar{\nu}}(\bm{R} \bm X_{0}) &=& h_{\mu\nu}(\bm X_{0})\\
    \label{eq:2e_rot}
    \pi_{\bar{\mu}\bar{\nu}\bar{\lambda}\bar{\sigma}}(\bm{R} \bm X_{0}) &=&\pi_{\mu\nu\lambda\sigma} (\bm X_{0})
\end{eqnarray}
Eqs. \ref{eq:1e_rot}-\ref{eq:2e_rot} are proven 
in Appendix \ref{app:ao_rot}. Here, if we rotate the molecule by a rotation $\bm{R}$, we also imagine rotating all of shells of basis functions around each atomic center by the corresponding unitary matrix $\hat{U}(\bm{X})$ so as to generate rotated basis functions. In  other words, if  $\ket{\nu} \in S$, for a specific shell (S), we define: $\ket{\bar{\nu}} = \sum_{\nu\in S}\ket{{\nu}} U_{\nu\bar{\nu}}$.

Lastly, as far as the $\bm \Gamma$ matrix is concerned, according to Eq. \ref{eq:Gamma_uv4prime}, we know that:
\begin{eqnarray}
\label{eq:Gamma_rot}
       \bm{R}\bP \cdot \bGamma^{A}_{\bar{\mu} \bar{\nu}}(\bm{R} \bm X_0) &=  \bm{R}\bP \cdot \bm{R}\bGamma^{A}_{\mu \nu}(\bm X_0) = \bP \cdot\bGamma^{A}_{\mu \nu}(\bm X_0)
\end{eqnarray}
Therefore, at the end of the day, all of the matrix elements within the energy expression $E_{PS}$ from Eq. \ref{eq:pssh_e} are rotationally invariant, and it follows from a variational treatement that the one and two electron density matrices must also be rotationally invariant ($D_{\bar{\mu} \bar{\nu}}(\bm{R} \bm X_0)=D_{\mu \nu}(\bm X_0)$, $G_{\bar{\mu} \bar{\nu}}(\bm{R} \bm X_0)=G_{\mu \nu}(\bm X_0)$ ). Thus, we may conclude that, so long as the  $\bm \Gamma$ matrix satisfies the four conditions given in Eqs. \ref{eq:Gamma_uv1}-\ref{eq:Gamma_uv4prime}, the energy $E_{PS}$ will also be rotationally invariant and satisfy Eq. \ref{eq:epsrot}.

\subsection{An Ansatz for $\Gamma $ From the Theory of Electron Translation and Rotation Factors \cite{Tian2023:ERF}}
For a practitioner of surface hopping dynamics,\cite{Tully1990,Barbatti2011,tretiak:2011:conj,Wang2016} one immediately recognizes the constraints in  Eqs. \ref{eq:Gamma_uv1}-\ref{eq:Gamma_uv4prime},  as these are the constraints that guide the construction of  electron translation factors (ETFs)\cite{Fatehi2011} and electron rotation factors (ERFs),\cite{athavale2023} which are discussed in a companion paper (Ref. \citenum{Tian2023:ERF}). See equations 15-16, 67, 85, in Ref. \citenum{Tian2023:ERF}. To that end, some words of background are appropriate here. 

During the course of a surface hopping trajectory, it is well established from model studies\cite{pechukas:1969,herman1984,Kapral1999,subotnik:2013:qcle_fssh_derive} that, when hopping from state $I$ to state $J$,  the nuclear momentum should be rescaled in the direction of the derivative coupling vector $\bm d_{IJ}$ in order to conserve energy (and establish a balance between kinetic and potential energy). However, from a mathematical point of view, this momentum rescaling scheme changes the total momentum because the derivative couplings satisfy
\begin{eqnarray}
\label{eq:d_trans}
    -i\hbar \sum_{A}d^{A\alpha}_{IJ}   &=&  -\bra{\Phi_{I}}\hat{p}^{\alpha}_{e}\ket{\Phi_{J}} \\
    \label{eq:d_rot}
     -i\hbar \sum_{A\beta\gamma}\epsilon_{\alpha\beta\gamma}X^{A\beta}d^{A\gamma}_{IJ} &=& -\bra{\Phi_{I}}\hat{L}^{\alpha}_{e}+\hat{s}^{\alpha}\ket{\Phi_{J}}  
\end{eqnarray}
The above equations for $\bm d_{IJ}$ are exactly the same as Eqs. \ref{eq:condition1} and  \ref{eq:condition3} for the $\bm \Gamma_{IJ}$ coupling and reflect a phase convention for the electronic states $\ket{\Phi_I}$ and $\ket{\Phi_J}$.\cite{Littlejohn2023} 
From a physical point of view, the RHS of Eqs. \ref{eq:d_rot} and \ref{eq:d_trans} are nonzero because whenever the nuclei move along one adiabatic state, that motion also drags around electrons.  Thus,  whenever the molecular nuclei are  translated and/or rotated, that motion  can  induce changes in electronic linear and/or angular momentum and eventually lead to  an electronic transition.

Now, the fact that rescaling along the derivative coupling direction destroyed the conservation of {\em nuclear} linear and angular  momentum has long bothered chemists\cite{Shu2020,athavale2023,Wu2023}.  For the most part, chemists have regarded this failure as a limitation of semiclassical mechanics and the desire has always been to fix this problem by restoring the linear and angular momentum of the nuclei in some fashion or another\cite{Fatehi2011,Shu2020,athavale2023,Tian2023:ERF}. To that end, in Refs. \citenum{Fatehi2011,athavale2023}, in order to restore nuclear momentum conservation, we previously sought to remove the offending translational and rotational component of the derivative coupling, i.e. which leads to the notion of ETFs and ERFs respectively.  

\subsubsection {Electronic Translation Factors} 
We begin with ETFs. ETFs restore linear momentum conservation and can be isolated by working in a translating basis\cite{Bates1958,Schneiderman1969,Thorson1978,Delos1981,Deumens1994,Errea1994,Illescas1998}. As a practical matter,  constructing an ETF to restore linear momentum conservation is quite trivial from an electronic structure point of view: in any AO basis, whether for multireference configuration interaction (MRCI)\cite{Lengsfield1984,Saxe1985,Lischka2004} or configuration interaction singles (CIS)\cite{Fatehi2011} or TDDFT (approximate) \cite{Chernyak2000,Send2010,Tavernelli2010, ou:2014:tddft_rpa,alguire:2014:tdhf, Ou2015, herbert:2014:jcp_dercouple}  wavefunctions, one always finds that the derivative coupling vector between states $I$ and $J$ naturally decomposes into:
\begin{align}
    \bm d^{tot}_{IJ} = \bm d^{ETF}_{IJ} + \tilde{\bm d}^0_{IJ} 
\end{align}
where  $\bm d^{ETF}_{IJ}$  is effectively  the matrix element for the one-electron momentum operator: 
\begin{align}
d^{ETF,A \alpha}_{IJ} &= \sum_{\mu \nu}  \left<I \middle| \Gamma'^
{A \alpha}_{\mu \nu} a_{\mu}^{\dagger} a_{\nu}  \middle|J\right> \\ 
\Gamma'^{A \alpha}_{\mu \nu}  &= \frac{1}{2i\hbar} p^{\alpha}_{\mu \nu} \left( \delta_{BA} + \delta_{CA}\right)\label{eq:etf}.
\end{align}
Here,  in Eq. \ref{eq:etf},  we assume $\ket{\mu}$ is centered on atom $B$ and $\ket{\nu}$ is centered on atom $C$.
 Mathematically, $\bm\Gamma'$ in Eq. \ref{eq:etf} clearly satisfies Eq. \ref{eq:Gamma_uv1} and \ref{eq:Gamma_uv2}.
 
On a very practical note, $\tilde{\bm d}^0_{IJ}$ is what remains after subtracting away the ETF component and it is straightforward to separate $\bm d^{ETF}_{IJ}$ from $\tilde{\bm d}^0_{IJ}$ because  the latter scales like $1/(E_I-E_J)$ and blows up around a conical intersection; the former does not.

\subsubsection{Electronic Rotation Factors}
Next we consider ERFs; altogether, a combination of ETFs and ERFs should restore linear and angular momentum. Now, while ERFs are rarely discussed in the literature (as compared with ETFs), by working in a basis that both translates and rotates, we have recently shown that one can indeed construct ERFs. \cite{athavale2023,Tian2023:ERF}
Thus, one can 
decompose the total derivative coupling into three parts: an ETF part, an ERF part of the derivative coupling vector, and the remaining part:
\begin{align}
    \bm d^{tot}_{IJ} = \bm d^{ETF}_{IJ} + \bm d^{ERF}_{IJ} + \bm d^0_{IJ}  
\end{align}
where 
\begin{align}
d^{ERF, A \alpha}_{IJ} &= \sum_{\mu \nu}  \left<\Phi_I \middle| \Gamma''^
{A \alpha}_{\mu \nu} a_{\mu}^{\dagger} a_{\nu}  \middle| \Phi_J \right> 
\end{align}
Here, the form of $\bGamma''$ is necessarily more complicated than the form for $\bGamma'$ because the latter can be a local one-electron operator and the former only a semi-local one-electron operator.\cite{athavale2023,Tian2023:ERF}  Using the same atom labels (A,B,C) as above for the ETFs (Eq. \ref{eq:etf}), according to Ref. \citenum{Tian2023:ERF}, one reasonable form for $\bGamma''$  is\cite{Tian2023:ERF}:
\begin{align}
    \bm{\Gamma}''^{A}_{\mu\nu} &= \zeta_{\mu\nu}^A\left(\bm{X}_A-\bm{X}_{\mu\nu}^0\right) \times \left(\bm{K}_{\mu\nu}^{-1}\bm{J}_{\mu\nu}\right),\label{eq:gamma_v1_final}
\end{align}
where 
\begin{align}    
     \bm{J}_{\mu\nu} &= \frac{1}{i\hbar}\braket{\mu}{\frac{1}{2}\left(\hat{\bm{L}}^{B}_{e}+\hat{\bm{L}}^{C}_{e}\right)}{\nu} .\label{eq:gamma_cond3}
\\
\zeta^A_{\mu\nu} &= \exp\left(-w \frac{2|(\bm{X}_A-\bm{X}_B)|^2 |(\bm{X}_A-\bm{X}_C)|^2}{|(\bm{X}_A-\bm{X}_B)|^2 + |(\bm{X}_A-\bm{X}_C)|^2}\right)\label{eq:zeta}\\
    \bm{X}_{\mu\nu}^0 &= \sum_A \zeta_{\mu\nu}^A\bm{X}_A /\sum_A \zeta_{\mu\nu}^A.\label{eq:center_munu} \\
    \bm{K}_{\mu\nu}&=-\sum_{A}\zeta_{\mu\nu}^A\left(\bm{X}_A-\bm{X}_{\mu\nu}^0\right)^\top\left(\bm{X}_A-\bm{X}_{\mu\nu}^0\right)\mathcal{I}_3 + \sum_A\zeta_{\mu\nu}^A\left(\bm{X}_A-\bm{X}_{\mu\nu}^0\right)\left(\bm{X}_A-\bm{X}_{\mu\nu}^0\right)^\top\label{eq:final_K}
\end{align}
Here $\mathcal{I}_3$ is a 3-by-3 identity matrix. The parameter $w$ here controls the locality and in Ref. \citenum{Tian2023:ERF} 0.3 ${\rm Bohr}^{-2}$ was found to be a safe choice. See the discussion below for the importance of locality for ERF.  The definition of one of the Cartesian components ($\eta$) of $\hat{{\bm L}}^{B}_{e}$ in Eq. \ref{eq:gamma_cond3} is
\begin{align}
    \hat{{L}}^{B\eta}_{e} &= \Big[({\hat{\bm r}_e}- \bm X_B) \times {\hat{\bm p}_e} \Big]_{\eta}=  \hat{L}^{\eta}_{e} - \sum_{\beta\gamma}\epsilon_{\eta\beta\gamma}X_{B\beta}\hat{p}^{\gamma}_{e},
\end{align}

Although slightly more involved, it can be proven\cite{Tian2023:ERF} that the one-electron matrix elements
\begin{align}
    \bm \bGamma_{\mu \nu} = 
    \bm \bGamma'_{\mu \nu} +
    \bm \bGamma''_{\mu \nu}
    \label{eq:Gamma:joe}
\end{align}
satisfy both Eq. \ref{eq:Gamma_uv1} and Eq.  \ref{eq:Gamma_uv3} above.
Finally, as shown in Ref. \citenum{Tian2023:ERF}, the ETF+ERF matrix $\Gamma_{\mu \nu}$ transforms correctly under translations and rotations of the molecule. In particular, Eqs. 67 and 85 in Ref. \citenum{Tian2023:ERF} are identical with Eqs. \ref{eq:Gamma_uv2} and \ref{eq:Gamma_uv4prime} above.

\subsubsection{A Phase-Space  Hamiltonian That Approximates Eq. \ref{Hshenvi}}
At the end of the day, choosing the $\bGamma$ couplings in Eq. \ref{HPS} to be the ETFs+ERFs from Ref. \cite{Tian2023:ERF} makes a great deal of sense. In short, rather than removing out the $\hat{\bm d}_{ETF}$ and $\hat{\bm d}_{ERF}$  vectors and rescaling along the $\hat{\bm d}_0$ direction  within a standard surface hopping calculation, we imagine including explicit momentum coupling to the $\hat{\bm d}_{ETF}$ and $\hat{\bm d}_{ERF}$ vectors within a phase-space electronic Hamiltonian. 
 Heuristically, this choice can be rationalized by rewriting Eq. \ref{Hshenvi} above as:
\begin{align}
  \label{Hshenvi2}
   \hH_{\rm Shenvi}(\bm X,\bP) &= \frac{\bP^2}{2\bM}   - i\hbar \frac{\bP \cdot \left( \hat{\bm d}_{ETF} + \hat{\bm d}_{ERF} + \hat{\bm d}_0 \right) }{\bM} - \hbar ^2 \frac{\hat{\bm d} \cdot \hat{\bm d} }{2 \bM} + \hE_{ad}(\bm X)
\end{align}
Our recommended choice of phase-space electronic Hamiltonian  (using Eqs. \ref{HPS}, \ref{eq:etf}, \ref{eq:gamma_v1_final} and \ref{eq:Gamma:joe} above) is equivalent to removing the $\bm P \cdot \hat{\bm d}_0$ term (which is gauge-dependent and explodes in the vicinity of avoided crossings but vanishes far from crossings) as well as the $\hat{\bm d} \cdot \hat{\bm d}$ term (which is often neglected in an  $\hbar$ expansion) in Eq. \ref{Hshenvi2}.

\section{The Missing Ingredient:  Electronic Momentum}
\label{sec:nafie}
We have made the hypothesis above that a meaningful choice for $\bGamma$ in Eq. \ref{HPS} is to set these couplings equal to the sum of the ETFs and ERFs in Eqs. \ref{eq:etf} and \ref{eq:gamma_v1_final}. One means of judging the value of such an ansatz for couplings is to compare the resulting predictions for electronic momentum  vis a vis Nafie's celebrated expression\cite{Nafie1983,Nafie1992,Nafie1997,Nafie2004} for electronic momentum.
At this point, a brief interlude (and review) of Nafie's work is appropriate. 

To begin our review, note that one obvious failure of the BO approximation is the fact that motion along an adiabatic surface does not carry any electronic momentum\cite{Diestler2012,Diestler2012QC,Diestler2013,Bredtmann2015}. Mathematically, for any electronic wavefunction $\ket{\Phi_I}$ constructed by Eq. \ref{eq:psi_BO}, $\braket{\Phi_I}{\hat{\bp}_e}{\Phi_I}=0$.
This failure can be addressed by going to higher order in BO theory.\cite{Deumens1994,Kutzelnigg2007,Nagashima2009,Okuyama2009,patchkovskii:2012:jcp:electronic_current,Diestler2013NBO,Gross2016,Engel2018,Engel2019,Engel2020}  and identifying either the momentum or the electronic flux.\cite{Engel2022, takatsuka:2021:jcp:flux_conservation,Pohl2016} Mathematically, if we regard the term $- i\hbar \frac{\bP \cdot \hat{\bm d}}{\bM}$ as the perturbation in Eq. \ref{Hshenvi}, one can construct the perturbed wavefunction as follows (by summing over all other electronic state $J$):

\begin{align}
\label{eq:psi}
    \ket{\Psi_I} = \ket{\Phi_I} - i\hbar \sum_{J\ne I}\frac{ \braket{\Phi_J}{\sum_A \frac{\bP^A \cdot \hat{\bd}^A}{M_A}  }{\Phi_I}}{E_I -E_J} \ket{\Phi_J}
\end{align}
where $\left< \Phi_I \middle| \hat{\bm d}^A \middle| \Phi_J \right> \equiv \bm d^A_{IJ}$.
We can now evaluate the expectation value of $\braket{\Psi_I}{\hat{\bp}_e}{\Psi_I}$:
\begin{align}
    \braket{\Psi_I}{\hat{\bp}_e}{\Psi_I}= 2 \hbar \text{Im}  \sum_{J\ne I} \frac{ \braket{\Phi_J}{\sum_A \frac{\bP^A \cdot \hat{\bd}^A}{M_A}  }{\Phi_I}}{E_I -E_J} \braket{\Phi_I}{\hat{\bp}_e}{\Phi_J}\label{eq:pe_nafie}
\end{align}
At this point, let us assume a complete basis, restrict ourselves to the case of no spin-orbit coupling, and use the commutator:
\begin{align}
    \left[ \hH_{e},\hat{\br}_e \right] = -i\hbar \frac{\hat{\bp}_e}{m_e},
\end{align}
From this exact relationship, it follows that 
\begin{align}
    (E_I - E_J) \braket{\Phi_I}{\hat{\bm r}_e}{\Phi_J}= -i\hbar \frac{1}{m_e} \braket{\Phi_I}{\hat{\bp}_e}{\Phi_J}\label{eq:rp_eqiv}
\end{align}
Then, if we plug Eq. \ref{eq:rp_eqiv} into Eq. \ref{eq:pe_nafie} and use the fact that $\bra{\Phi_{I}} \hat{\bd}^A\ket{\Phi_{I}}= 0$, we find:
\begin{align}
    \braket{\Psi_I}{\hat{\bp}_e}{\Psi_I} &= 2 m_e  \text{Re}  \sum_{J } \braket{\Phi_J}{\sum_A \frac{  \bP^A \cdot \hat{\bd}^A}{M_A}  }{\Phi_I}\braket{\Phi_I}{\hat{\br}_e}{\Phi_J} \\
    & = 2 m_e  \text{Re}   \braket{\Phi_I}{ \hat{\br}_e \sum_A \frac{  \bP^A \cdot \hat{\bd}^A}{M_A}  }{\Phi_I} \\
     & = 2 m_e  \text{Re}   \braket{\Phi_I} {\hat{\br}_e }{\frac{d}{dt}\Phi_I} \\
      \braket{\Psi_I}{\hat{\bp}_e}{\Psi_I} & = m_e  \frac{d}{dt}  \braket{\Phi_I} {\hat{\br}_e}{\Phi_I}\label{eq:nafie}
\end{align}

Eq. \ref{eq:nafie} is Nafie's celebrated final result for the electronic momentum.
Clearly, Eq. \ref{eq:nafie} should be satisfied (or approximately satisfied) with a meaningful choice of $\bGamma$. In the extremely well-separated adiabatic limit, where there are minimal nonadiabatic interactions, this expression is the physical electronic momentum.
One means of checking the validity of any choice of $\bGamma_{\mu \nu}$ (with the corresponding phase-space Hamiltonian in Eq. \ref{HPS}) is to evaluate the electronic momentum and see whether or not the result satisfies Eq. \ref{eq:nafie}.

As a side note, we mention that Eq.\ref{eq:nafie} is valid in the  adiabatic limit, i.e. when moving along a well-separated adiabat $E_I$.  If one performs a nonadiabatic simulation and wishes to estimate the electronic momentum from a collection of states using a complete basis and a full CI Hamiltonian, one can derive a proper expression for the electronic momentum by using the corresponding electronic density matrix; as shown by Takatsuka\cite{takatsuka:2021:jcp:flux_conservation},  one can even check the validity of the continuity equation for the electronic momentum probability density in a finite basis.  One hope for the present electronic phase-space Hamiltonian is that one will be able to compute electronic momenta from a single adiabatic state calculation (as opposed to a full nonadiabatic calculation) in the adiabatic limit.

\subsection{Beyond Nafie: The Search For Electronic Angular Momentum}

Beyond linear momentum, we would also very much like a means to benchmark the predicted electronic angular momentum as well.  Unfortunately, however, the approach from the previous section cannot be generalized to the case of angular momentum and there is no simple final expression (as in Eq. \ref{eq:nafie}) for  $\braket{\Psi_I}{\hat{\bm L}_e}{\Psi_I}$. The reason is simple: While Eq. \ref{eq:nafie} above is the quantum mechnanical analogue of the classical expression, $\bm p_{e} = m_{e} \frac{d \bm r_{e}}{dt},$ there is no analogous expression for rotations. For a simple rotation in two dimensions, one can write 
$  L = { \mathcal{I}} \frac{d \theta}{dt}$, where $ L$  is the angular momentum, $ { \mathcal{I}}$ is the moment of inertia, and $\theta$ is the relevant angle, but for a  multidimensional problem, there is no unique $\theta$ and ${ \mathcal{I}}$ becomes a matrix.  Mathematically, one big issue is commutativity: whereas $\hp^x_{e}$ and  $\hp^y_{e}$ commute, $\hL^x_{e}$ and $\hL^y_{e}$ do not commute, and this lack of commutativity prevents any multidimensional analogue of 
 Eq. \ref{eq:nafie} for angular momentum. Thus, in principle, if one wishes to estimate the electronic angular momentum,  one needs to evaluate a full sum over states as in  Eq. \ref{eq:pe_nafie}, which is indeed painful.

Nevertheless, in the limit of the  two dimensional motion of a rigid linear (or linear like) molecule in the $xy$ plane, the algorithm above can  be partially adapted by recognizing that the relevant quantum mechanical expression is:
\begin{eqnarray}
    \hH = -\frac{\hbar^2}{2m_{e}r_e^2}\frac{\partial^2}{\partial \theta^2} = \frac{\hL_e^2}{2m_{e}r_e^2}\\
     \left[ \hH, m_{e} r_{e}^2 \hat{\theta} \right] = -i\hbar \cdot \hat{L}_e
\end{eqnarray}
Here we transform to relative coordinates, assume the center of mass is placed at the origin, and define $r_e^{2} = x^{2} + y^{2}$ as the distance of the particle away from the origin.

Proceeding as we did above between Eqs. \ref{eq:pe_nafie} and \ref{eq:nafie}, it follows that:
\begin{eqnarray}
 \label{eq:Le_fd}
 \braket{\Psi_I}{\hat{L}_{e}}{\Psi_I} & = m_{e}  \frac{d}{dt}  \braket{\Phi_I}{r_e^2 \hat{\theta} }{\Phi_I}   = m_e r_e^2\frac{d}{dt} \int  \theta \left| \Phi_I(\theta) \right|^2 d \theta
\end{eqnarray}
For a linear or nearly linear molecule aligned along the x-axis, let us  make the assumption that the electronic density is localized around $\theta = 0$ and $\theta = \pi$ (but vanishes around $\theta = \pm \pi/2$). In such a case,  we can expand the Cartesian coordinates $(x,y) = (r_{e}cos(\theta),r_{e}sin(\theta))$ around $\theta = 0$ and $\theta = \pi$,  and obtain $(x,y)|_{\theta=0} \approx (r_{e},r_{e}\theta) $ and $(x,y)|_{\theta=\pi} \approx (-r_{e},-r_{e}(\theta-\pi)) $ at the zeroth order, respectively. 
Therefore, in the vicinity of $\theta =0$, $xy = r_{e}^2\theta$; in the vicinity of $\theta = \pi$, $xy = r_{e}^2\theta-r_{e}^2 \pi.$
Note that because the integral $\bra{\Phi_I}-\pi r^2_{e} \ket{\Phi_I}$ is a constant (for a rigid rotor with $r_e$ fixed), this terms gives zero when taking the time derivative.
Therefore,  we can evaluate the  final integral as:
\begin{eqnarray}
     m_{e}  \frac{d}{dt}  \braket{\Phi_I}{r_e^2 \hat{\theta} }{\Phi_I}  =  m_{e}  \frac{d}{dt}  \braket{\Phi_I}{\hat{x} \hat{y}}{\Phi_I} = -   \frac{d}{dt}  {\mathcal{I}_{xy} }\label{eq:nafieL}
\end{eqnarray}
Below, we will check whether our electronic phase-space Hamiltonian satsifies Eq. \ref{eq:nafieL} for the case of linear  molecules undergoing rigid rotation.

\section{Numerical Results}\label{sec:result}

We will now present numerical results  testing the ability of our proposed phase-space electronic Hamiltonian in Eq. \ref{HPS} to recover the correct linear and angular momentum. Our numerical results are within a Hartree-Fock (HF) framework which, though not exact, should offer a good enough starting point. The benchmarks for linear and angular momentum were computed by a finite-difference (FD) Hartree-Fock approach according to Eq. \ref{eq:nafie} and Eq. \ref{eq:Le_fd}, respectively. All the geometries used were optimized at the HF level with a  cc-pVTZ basis set and the coordinates are given in Appendix \ref{app:coord}. We investigate the effects of different basis sets, and use the acronyms XZ for cc-pVXZ and aXZ for aug-cc-pVXZ basis sets in Tables \ref{tab:translation}-\ref{tab:rotation_gamma} below. All the linear molecules are aligned along the x-axis.

We begin by checking the accuracy of the linear electronic momentum as computed with the phase-space Hamiltonian when the entire molecule translates.  In Table \ref{tab:translation}, we show the $\alpha$ components of the linear momentum as calculated with a translation velocity along the $\alpha$ axis and a magnitude corresponding to room temperature thermal motion. The FD approach uses a time step of 1 a.u. (1 a.u. = 0.0242 fs).  The linear momentum values calculated with the FD approach (Eq. \ref{eq:nafie}) serves as the benchmark and does not change much at all with different basis sets for translation motion. According to Table \ref{tab:translation}, the linear momentum calculated with the phase-space Hamiltonian approaches do converge to the finite-difference values for larger basis sets. This convergence demonstrates that, despite using a single state, the practical phase-space method at the Hartree-Fock level can effectively capture the electronic  motion going slightly beyond the BO approximation--when given a reasonably large basis set. Note that adding the one-electron rotation factor $\bm\Gamma''$ does not change the results in Table \ref{tab:translation} at all due to the invariance of $\bm\Gamma''$ with respect to translation, $\sum_{A}\Gamma''^{A\alpha}_{\mu\nu} = 0$.
\begin{table}[H]
\centering
\caption{The $\alpha$ components of the linear momentum ($\hbar/a_{0}$) calculated with a translational velocity along the $\alpha$ axis. The linear molecules are aligned along the $x$ axis, and  we report translations for the $\alpha = x,y$ directions. The finite-difference (FD) approach in Eq. \ref{eq:nafie} serves as the benchmark and does not change with different basis sets. The linear momentum calculated with the phase-space Hamiltonian approaches the finite-difference values for larger basis sets.}
\label{tab:translation}
\renewcommand{\arraystretch}{2}
\begin{tabular}{cccccccccc}
\hline
 & & STO-3G &DZ & aDZ & TZ & aTZ & QZ & aQZ & FD \\ \hline
 \ce{H_{2}}& <$\hat{p}_e^{x}$>& 7.94e-4 &1.41e-3&1.41e-3&1.42e-3&1.43e-3&1.43e-3 & 1.43e-3 & \multirow{2}{*}{1.43e-3}  \\
 & <$\hat{p}_e^{y}$>   & 0.00 & 7.69e-4 & 1.41e-3 &1.20e-3 &1.43e-3  &1.33e-3  &1.43e-3 \\\hline
 \ce{LiH}& <$\hat{p}_e^{x}$> & 2.83e-4 & 7.77e-4 &9.07e-4&1.10e-3&1.16e-3&1.28e-3&1.31e-3 & \multirow{2}{*}{1.44e-3} \\
 & <$\hat{p}_e^{y}$> & 2.22e-4&6.07e-4&7.51e-4&9.14e-4&1.01e-3&1.21e-3&1.25e-3 \\\hline
  \ce{HCN} & <$\hat{p}_e^{x}$> & 1.18e-3 & 2.09e-3&2.16e-3&2.39e-3&2.43e-3&2.62e-3&2.63e-3 & \multirow{2}{*}{2.74e-3}\\ 
  & <$\hat{p}_e^{y}$> & 1.80e-4 & 1.63e-3&2.00e-3&2.25e-3&2.38e-3&2.57e-3&2.61e-3\\\hline
 \ce{H_{2}O}$^{\dagger}$ & <$\hat{p}_e^{x}$> & 5.06e-4&1.63e-3&1.93e-3&2.09e-3&2.18e-3&2.28e-3&2.32e-3&\\
  & <$\hat{p}_e^{y}$> & 4.47e-4 &1.60e-3&1.92e-3&2.08e-3&2.18e-3&2.28e-3&2.32e-3&2.40e-3\\
  & <$\hat{p}_e^{z}$> & 2.03e-5 & 1.39e-3&1.86e-3&2.01e-3&2.16e-3&2.24e-3&2.31e-3 &\\\hline
\end{tabular}
\justifying 
$^{\dagger} $ For the triatomic water molecule, we report the electronic momentum in the $x-$direction--assuming we translate the molecule in the $x-$direction (so forth for the $y$ and $z$ directions).  For the FD approach, assuming we translate along the $x-$direction,  we find <$\hat{p}_e^{y}$> = <$\hat{p}_e^{z}$> = 0 and so all relevant information is included in Table \ref{tab:translation}. For the phase-space approach, $\left<\hat{p}_e^{y}\right>$ can be non-zero, but the momentum is usually at least two orders of magnitude smaller than $\left<\hat{p}_e^{x}\right>$, and so is not included  in the table either.
\end{table}

Next we examine the linear momentum  for internal motion. In Table \ref{tab:vibration}, we report the linear momentum when moving only the H atom with a velocity that again corresponds to room temperature thermal motion. According to Table \ref{tab:vibration}, the linear momentum calculated with the FD approach does converge with larger basis sets. By contrast, for molecules other than the \ce{H_{2}} molecule, the linear momentum as calculated with our phase-space approach does not converge as well with larger basis sets, insofar as diffuse functions clearly creates differences. Nevertheless, when a decently sized non-augmented basis set is used, such as a cc-pVDZ basis set, we find the linear momentum calculated with the two approaches do agree at least qualitatively. Note that here the one-electron rotational factor $\bm \Gamma''$ has not been included. Including the rotational factor  changes the results of \ce{H_{2}O} molecule by only small amount, as shown in Table \ref{tab:h2o_gamma} in the Appendix \ref{app:gamma_contributions}. 

The discrepancies between the two approaches in Table \ref{tab:vibration} can be understood as follows for the specific case of a HF ansatz. Let us write out Nafie's expression for the electronic momentum in an AO basis (where $D_{\mu \nu}$ is the relevant density matrix):
\begin{align}
    \left<\hat{\bm p}_e\right> &= m_e \frac{d}{dt}\left<\hat{\bm r}_{e}\right> \\ 
        &= m_e  \frac{d}{dt} \sum_{ \mu \nu} D_{\mu \nu} \bm r_{\mu \nu} \\
        &= m_e \sum_{\mu \nu} \left( \frac{d}{dt} D_{\mu \nu} \cdot  \bm r_{\mu \nu} + D_{\mu \nu} \cdot \frac{d}{dt} \bm r_{\mu \nu} \right)
\end{align}
The first term accounts for the fact that the density matrix (in a moving AO basis) changes as a function of time (so-called orbital response); the second term accounts for the fact that the AO basis itself changes as a function of time. 
For the ideal case of an H-atom with the electron in a 1s orbital, the first term would be zero while the second term would be nonzero. In view of this analysis, it is is clear that our proposed phase-space electronic Hamiltonian captures all of the physics in the second term but not  the physics in the first term (to lowest order). In practice,  whenever the systems passes through a crossing with another state, the $\frac{d}{dt}D_{\mu \nu}$ term will get larger and we will not capture such nonadiabatic effects.

\begin{table}[H]
\centering
\caption{The nonzero components of the linear momentum ($\hbar/a_{0}$) calculated with a velocity that corresponds to stretching one H atom along the $x$ axis. All the linear molecules are aligned along the $x$ axis. The finite-difference (FD) approach serves as the benchmark and the results converge with larger basis sets. The linear momentum calculated with the phase-space Hamiltonian matches qualitatively the finite-difference values for reasonably large basis sets.}
\label{tab:vibration}
\renewcommand{\arraystretch}{2}
\begin{tabular}{ccccccccc}
\hline
 & & STO-3G &DZ & aDZ & TZ & aTZ & QZ & aQZ  \\ \hline
 \ce{H_{2}} & <$\hat{p}_e^{x}$>  & 5.62e-4 & 9.96e-4 & 1.00e-3 & 1.00e-3 &1.01e-3 & 1.01e-3 &1.01e-3 \\
  & FD  & 1.01e-3 &1.01e-3 &1.01e-3&1.01e-3 &1.01e-3& 1.01e-3& 1.01e-3 \\\hline
 \ce{LiH} & <$\hat{p}_e^{x}$> &1.87e-4 & 7.17e-4&9.62e-4&5.68e-4&4.26e-4&4.48e-4&2.01e-4\\
 & FD & 1.07e-3 & 1.45e-3 & 1.48e-3 &1.50e-3 & 1.51e-3 &1.50e-3& 1.51e-3 \\\hline
 \ce{HCN} & <$\hat{p}_e^{x}$> & 5.30e-4&6.76e-4&9.54e-4&1.02e-3&7.09e-4&1.14e-3&7.34e-4 \\
 & FD & 6.16e-4&7.16e-4&7.24e-4&7.34e-4&7.25e-4&7.27e-4&7.26e-4\\\hline
 \ce{H_{2}O}$^{*}$ & <$\hat{p}_e^{x}$> & 3.57e-4 & 7.11e-4 & 6.67e-4 & 2.75e-4 &5.13e-4&2.03e-4&7.74e-4 \\
 & FD & 1.17e-3&8.16e-4&8.06e-4&8.19e-4&8.02e-4&8.09e-4&8.02e-4\\
 & <$\hat{p}_e^{y}$> & 5.82e-6&-6.86e-5&-4.20e-5&-1.08e-4&-1.21e-5&-1.96e-4&-7.21e-5\\
 & FD & -4.59e-5 & -1.96e-5 &-5.07e-5&-3.98e-5&-5.01e-5&-4.57e-5&-5.05e-5\\\hline
\end{tabular} \\
$^{*}$ The water molecule is placed in the xy plane with one of the O-H bonds aligned along the x-axis. Stretching this O-H bond along the x-axis yields a relatively large linear momentum in the x direction ($\left<\hat{p}_e^{x}\right>$) and a relatively small linear momentum in the y direction ($\left<\hat{p}_e^{y}\right>$).
\end{table}

\begin{figure}[H]
\includegraphics[width=0.7\textwidth]{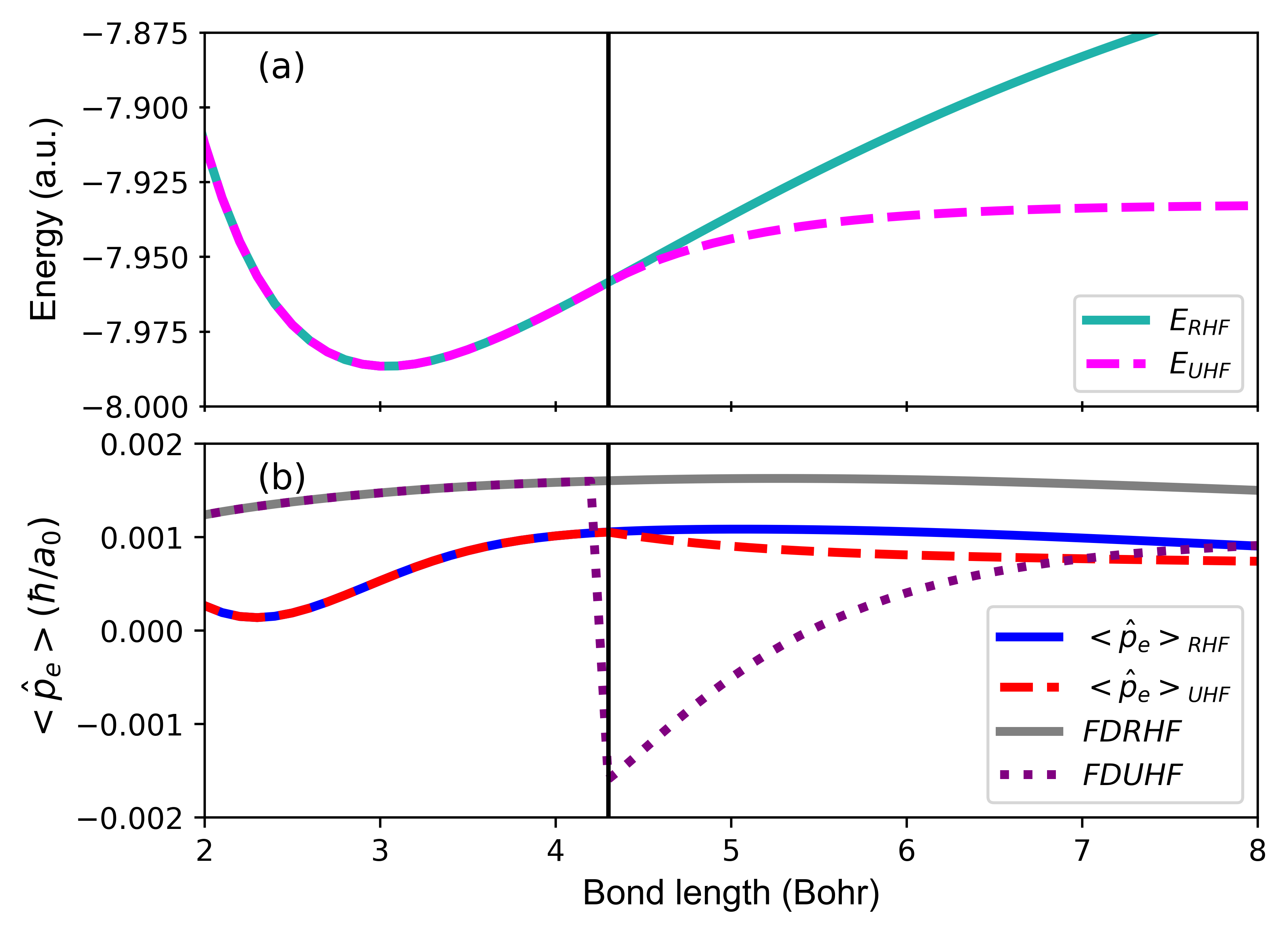}
\caption{ (a) RHF and UHF potential energy surfaces drawn with a cyan solid line and a magenta dashed line, respectively. (b)The linear momentum calculated with the phase-space restricted HF (blue solid line), phase-space unrestricted HF (red dashed line), the FD restricted HF (grey solid line), and FD unrestricted (purple dotted line). The equilibrium bond distance is 3.0374 Bohr. The location of the Coulson-Fischer point is shown by the black solid line, where the internal nuclear motion  changes the electronic wavefunction discontinuously so that the FD approach predicts a spike in the linear momentum. The phase-space HF approach does not take into account the orbital response to the nuclear motion and yields a smoother prediction in this region. In this figure, we have set $P_{H} = 1.8 $ a.u. (i.e. $v_{H} = 0.001$ a.u.).}\label{fig:lih}
\end{figure}

In order to further probe how much of the electronic momentum depends on nonadiabaticity, we have
investigated the electronic momentum as a function of bond distance for the case of LiH molecule, a molecule where curve crossings are known to occur at large separation. In Figure \ref{fig:lih}, we plot the linear momentum calculated with the phase-space restricted HF (blue solid line) and unrestricted HF (red dashed line) approaches, and with the FD approach with restricted HF (grey solid line) and unrestricted HF (purple dotted line). The equilibrium bond distance is 3.0374 Bohr and we set the hydrogen momentum to be $P_H = 1.8 $ a.u. (i.e. $v_H = 0.001$ a.u.); here a positive value of $P_H$ stretches the molecule. 
Between 2 and 8 Bohr, there is a Coulson-Fischer (CF) point at $\sim$4.3 Bohr,
where the HF state transitions from closed shell (ionic) to open shell (biradical).
Around this CF point, the FD unrestricted approach yields a negative spike in linear momentum.
Physically, when we increase the bond distance,
there is a large decrease in the electronic position expectation value  $\bra{\Phi_{I}}\hat{x}\ket{\Phi_{I}}$ because, as the molecule transitions from the ionic configuration (Li$^{\delta +}$--H$^{\delta -}$)
to a biradical configuration (Li$^{\bullet}$--H$^{\bullet}$), and electron jumps back from the H nucleus to the Li nucleus.
At this CF point, where the orbital response is enormous, obviously the two expressions for electron linear momentum (phase-space vs finite difference) strongly disagree--but otherwise they are the same order of magnitude.

Next, let us compare the angular momentum calculated with the phase-space approach and the finite difference approach according to Eq. \ref{eq:Le_fd}. In Table \ref{tab:rotation}, we show the angular momentum computed with both approaches for molecules at both the equilibrium and  stretched geometries. The bond lengths for the stretched geometry were 8 Bohr. All of the molecules are placed with their center of mass at the origin and are aligned along the $x-$axis. A velocity corresponding to a rigid rotation around the $z-$ axis was applied. For the diatomic molecules, this velocity corresponded to a rotation of the whole molecule by 0.05 degrees per time step (1 a.u.). For the polyatomic molecules, a smaller velocity was chosen such that the total rotational kinetic energy corresponded to room-temperature thermal motion; as above, for the FD calculations, we set the time step to be 1 a.u.. For the equilibrium geometries, we find the angular momentum calculated with the phase-space and the finite-difference approaches are about the same magnitude for reasonably sized basis sets. This agreement improves for the stretched geometry, which makes sense given the assumptions in Eq. \ref{eq:Le_fd}. After all, we assumed there that the angle of the electronic density in the $xy$ plane was centered mostly around $\theta=0$ and $\theta=\pi$ and if we compare stretched versus equilibrium geometries,  we find that the electronic density  centered around the origin is much less in the former (rather than latter) geometries.

\begin{table}[H]
\centering
\caption{The $z$ components of the angular momentum ($\hbar$) calculated with a velocity that corresponds to a rigid rotation around the $z$ axis. All the linear molecules were aligned along the $x$ axis with the center of mass placed at the origin. The finite-difference (FD) approach serves as the benchmark and converges quickly with larger basis sets. The angular momentum calculated with the phase-space Hamiltonian agrees reasonably well with the finite-difference values given a decent size of basis sets. The agreement improves at the stretched geometry of \ce{H_{2}} as the majority of the densities distribution reside within a small angle with respect to the x axis.}
\label{tab:rotation}
\renewcommand{\arraystretch}{2}
\begin{tabular}{ccccccccc}
\hline
 & & STO-3G &DZ & aDZ & TZ & aTZ & QZ & aQZ  \\ \hline
 \ce{H_{2}} & <$\hat{L}_e^{z}$> & 0.00 & 5.28e-5 & 7.50e-5 & 7.21e-5 &7.28e-5& 7.29e-5 &7.35e-5 \\
  & FD & 5.71e-4 & 4.58e-4 & 4.53e-4 &4.42e-4 & 4.11e-4 &4.25e-4& 4.15e-4  \\\hline
  Stretched \ce{H_{2}} & <$\hat{L}_e^{z}$> &  0.00 & 6.91e-3 & 2.26e-2 & 1.34e-2 &2.36e-2&1.70e-2&2.38e-2\\
  &FD & 4.84e-3&2.76e-2 & 2.61e-2 &2.72e-2&2.60e-2&2.69e-2&2.59e-2\\\hline
 \ce{LiH} & <$\hat{L}_e^{z}$> &5.74e-3 & 9.38e-3&1.07e-2&1.07e-2&1.12e-2&1.11e-2&1.13e-2\\
 & FD & 7.60e-3 & 8.98e-3 & 9.27e-3 &9.37e-3 & 9.50e-3 &9.46e-3& 9.51e-3 \\\hline
 \ce{HCN} & <$\hat{L}_e^{z}$> &2.47e-3& 3.49e-3&3.70e-3&3.98e-3&4.05e-3&4.25e-3&4.27e-3\\
 & FD & 3.80e-3 &3.62e-3&3.60e-3&3.60e-3&3.59e-3&3.60e-3&3.59e-3\\\hline
 \ce{C_{4}H_{2}} & <$\hat{L}_e^{z}$> & 1.54e-3 & 7.49e-3&8.88e-3&9.71e-3&1.03e-2&1.09e-2&1.11e-2 \\
 & FD & 1.14e-2&1.13e-2&1.12e-2&1.13e-2&1.13e-2&1.13e-2&1.13e-2\\\hline
 \ce{}
\end{tabular}
\end{table}

\section{Discussion: Semi-Locality of $\bm \Gamma''$}
As discussed above (as well as in Ref. \citenum{athavale2023} and the companion paper [Ref. \citenum{Tian2023:ERF}]),  the ERF term $\bm \Gamma''$ cannot be made strictly local -- unlike the ETF term $\bm \Gamma'$. To that end, in Eq. \ref{eq:zeta}, we introduced a parameter $w$ to control the locality of $\bm \Gamma''$. Now, the results in Table \ref{tab:vibration} and \ref{tab:rotation} are clearly not very sensitive to the choice of parameter $w$.
One might wonder: does this insensitively imply that $w$ is not of practical importance? Or is the insensitivity perhaps just a consequence of the fact that we modeled 
small molecules exclusively in the tables above. To prove that the former is false and the latter is correct, i.e. that   semi-locality of $\bm \Gamma''$ is indeed crucial for a physical meaningful effect,  in this section, we will work with simple numerical example
illustrating how and why maintaining size-consistency is essential.  

Let us consider one rigidly rotating \ce{LiH} molecule (studied in Table \ref{tab:rotation}) at the origin and add a second \ce{LiH} molecule around $\sim$35 Bohr away. (See Appendix \ref{app:coord} for the geometry of this \ce{LiH} dimer). One expects that the second \ce{LiH} molecule, located far away, should not affect the angular momentum resulted from the rigid rotation of the original \ce{LiH} molecule. In Table \ref{tab:2lih_rot}, we list <$\hat{L}_e^{z}$> values for the original \ce{LiH} molecule as calculated either when the molecule is isolated (same data as from Table \ref{tab:rotation} and Table \ref{tab:rotation_gamma}) or in the presence of second  \ce{LiH} molecule. We present data from both    $w=0$  (corresponding to no locality) and $w=0.3$ (corresponding to semi-locality) calculations.  Notice that the results calculated with semi-locality ($w=0.3$) match with the <$\hat{L}_e^{z}$> calculated for the single \ce{LiH} system, indicating that the second \ce{LiH} molecule far away correctly has negligible impact on the <$\hat{L}_e^{z}$> values. By contrast, without any locality constraint ($w=0$), the   <$\hat{L}_e^{z}$> values are different (incorrectly).  

As a side note, in Table \ref{tab:2lih_rot}, we also list the <$\hat{L}_e^{z}$> values as calculated with only the ETF term $\bm\Gamma =\bm\Gamma'$. Here, the single and double  \ce{LiH} systems yield the same exact expectation values as one should expect from size consistency -- again because  $\bm\Gamma'$ is strictly local. That being said, in Table \ref{tab:2lih_rot}, we find that the results with $w=0$ for the double \ce{LiH} 
 case  are very similar to those with $\bm\Gamma'$ alone; clearly,  the $\bm\Gamma''$ contribution to <$\hat{L}_e^{z}$> values becomes too small if one does not enforce locality on the ERF  matrix elements. 

\begin{table}[H]
\centering
\caption{<$\hat{L}_e^{z}$>  values ($\hbar$) as calculated for a rigid rotation of one \ce{LiH} molecule in the absence and in the presence of another molecule located far away in space.  Because of the large distance between the molecules, the angular momentum of the first \ce{LiH} molecule should be the same in both calculations. The phase-space approach results are reported with $\bm\Gamma =\bm\Gamma'$ only and $\bm\Gamma = \bm\Gamma' + \bm\Gamma''$. When $\bm\Gamma''$ is included, two choices of the locality parameter $w=0$ (non-local) and $w=0.3$ (semi-local) are investigated for the two  \ce{LiH} system. (For the single \ce{LiH} system, $w=0$ and $w=0.3$ give the same results.) Only the $w = 0.3$ data gives consistent results for the \ce{LiH} molecule.  The data here proves conclusively that one cannot pick $w = 0$ (and ignore locality) if one seeks meaningful results. }
\label{tab:2lih_rot}
\renewcommand{\arraystretch}{2}
\begin{tabular}{ccccccccc}
\hline
  & & STO-3G &DZ & aDZ & TZ & aTZ & QZ & aQZ  \\ \hline
Two \ce{LiH}  & $\bm \Gamma'$ & 9.11e-4&3.50e-3&6.22e-3&4.93e-3&7.55e-3&5.77e-3&6.85e-3 \\
 & $\bm\Gamma' + \bm\Gamma'' (w=0)$ & 6.27e-4&3.15e-3&5.96e-3&4.59e-3&7.33e-3&5.46e-3&6.58e-3 \\
& $\bm\Gamma' + \bm\Gamma'' (w=0.3)$ &5.74e-3 & 9.38e-3&1.07e-2&1.07e-2&1.12e-2&1.11e-2&1.13e-2\\\hline
 Single \ce{LiH}  & $\bm\Gamma'$  & 9.11e-4&3.50e-3&6.22e-3&4.93e-3&7.55e-3&5.77e-3&6.85e-3 \\
 & $\bm\Gamma' + \bm\Gamma''$ &5.74e-3 & 9.38e-3&1.07e-2&1.07e-2&1.12e-2&1.11e-2&1.13e-2\\
 & FD & 7.60e-3 & 8.98e-3 & 9.27e-3 &9.37e-3 & 9.50e-3 &9.46e-3& 9.51e-3 \\\hline
\end{tabular}
\end{table}

\section{Discussion: Implications for Nonadiabatic Dynamics}\label{sec:discussion}
The theory and results above present a reasonably compelling argument that, in the future, one can improve upon Born-Oppenheimer dynamics simply by working with a phase-space electronic Hamiltonian that depends on both nuclear position and momentum. 
Admittedly, the data gathered here is still limited. We have checked for the electronic linear momentum with regards to internal motion, but we have not yet checked for the electronic angular momentum with regards to internal motion (only with regards to rigid motion). 
Checking for internal motion will require a large basis and an exact full configuration interaction (FCI) calculation in order to evaluate the exact angular momentum through the exact sum over states expansion in Eq. \ref{eq:psi}. Such a daunting task will probably be necessary in the future.
Nevertheless, notwithstanding its limited nature, the data gathered so far is quite encouraging and has potential to open up new areas of study\cite{Nakamura1980}.

Obviously, launching dynamical simulations in the near future will be the next logical step.
To that end, note first that
the $\bGamma$ couplings are one-electron operators such that diagonalizing the resulting phase-space electronic Hamiltonian should require a trivial added cost. 
The same cost analysis holds for the gradients of this phase-space electronic Hamiltonian, such that running {\em ab initio} dynamics should be readily possible. Moreover,  
one can imagine running both single surface dynamics (analogous to BO dynamics) as well as nonadiabatic surface hopping dynamics. In the spirt of Ref. \citenum{Wu2023}, we can be confident that all such dynamics will both predict nonzero electronic angular momentum while also conserving the total angular momentum. To date, the only semiclassical nonadiabatic algorithm for simulating such dynamics has been through the exact factorization approach\cite{gross:2022:prl:angmom} (which is another promising approach but still under development\cite{Abedi2010,Abedi2012,gross:2016:exact_factorization}).

Another obvious target of the current research is the calculation of vibrational circular dichroism (VCD) spectra.\cite{Stephens1985Review,Nafie1997VCDreview,Vass2011,Nafie2020} Indeed, the inability of standard BO theory to calculate VCD spectra stimulated the original work of Nafie,\cite{Nafie1977,Freedman1983,Nafie1983,Nafie1992} and now many others,\cite{Stephens1985,Stephens1987,Rauk1992,Vuilleumier2015} to go beyond BO theory and construct the relevant matrix elements (with electronic angular momentum along the ground state) that allowed for a nonzero vibrational rotatory strength.  While the present phase-space approach has many similarities with the nuclear velocity perturbation (NVP) approach in Refs. \citenum{Nafie1992,Sebastiani2013,Sebastiani2016,Luber2022}, it is worth noting that with a phase-space electronic Hamiltonian, apparently one can calculate a nonzero rotatory strength without doing a double response theory (albeit at the cost of ignoring orbital response).  Thus, the present approach would appear to be a natural starting point for VCD calculations in the future.  \cite{joefootnote} 

Although not addressed above, it is crucial to emphasize that all of the theory above generalized to include spin degrees of freedom. In such a case, as discussed in the Conclusion and Outlook section of Ref. \citenum{Tian2023:ERF}, one needs only modify the form of $\bm\Gamma''$ to include spin degrees of freedom and the total angular momentum is conserved (without explicitly including a Berry force). In recent years, there has been an explosion of interest in coupled nuclear-electronic-spin motion, highlighted by the possibility that nuclear motion (and chiral phonons) may yield new insight into the chiral-induced spin selectivity (CISS) effect,\cite{Naaman2012,Naaman2019,Evers2022,Das2022,Fransson2023} whereby electronic motion through molecular sytems shows clear signs of spin preference.  Beyond CISS, one can also imagine running similar dynamics to explore spin polarization in intersystem crossing more generally.\cite{Hall1968,El‐Sayed1970,Hall1971,Hall1971:MP,Hall1975} 

Finally, beyond dynamics, one can envision that because the chemistry and physics communities are less experienced with phase-space Hamiltonians than with standard electronic Hamiltonians, new tools will be needed in the future. 
From a bird's eye view, as far as the electrons are concerned, a phase-space electronic Hamiltonian breaks time-reversal symmetry and is equivalent to introducing a fluctuating magnetic field, and so advanced statistical mechanics sampling methods will be needed. And if there are spin degrees of freedom, these statistical mechanics methods will need to be compatible with closely spaced, nearly or fully degenerate electronic states.  Lastly, the problems will be only richer if we include explicit magnetic fields\cite{Steiner1989,Coey2010,Helgaker2012} as well.

At the end of the day, once the computational tools have been built, there is the potential to use the current phase-space approach so as to generalize the well-known Marcus parabolas\cite{marcus:sutin:1985,nitzanbook} to include spin degrees of freedom and to study how spin affects electron transfer and curve crossings in a manner which conserves angular momentum.  This approach should allow us to  explore very new physics involving the flow of angular momentum between nuclear, electronic and spin degrees of freedom.

\section{Summary}
In this paper, we have proposed a phase-space electronic Hamiltonian $H_{PS}(\bm X,\bm P)$ with an effective one-electron operator $\bm \Gamma_{\mu \nu}$ that couples electronic motion to the nuclear momentum $\bm P$.  
Our ansatz is that one can build the $\bGamma$ couplings using previously derived electron-translation and electron-rotation factor. 
These matrix elements satisfy Eqs. \ref{eq:Gamma_uv1}- \ref{eq:Gamma_uv4prime}, so that  
by including the $\bGamma \cdot \bm P$ term,  one naturally conserves the total linear and angular momentum, allowing the electronic and nuclear degrees of freedom to exchange linear and angular momentum.
Moreover,  our initial data suggests that  for this choice of $\bm \Gamma$,
one can qualitatively recover the correct electronic linear momentum in agreement with  Nafie's theory, as well as the correct electronic angular momentum (albeit for the case of rigid motion)--which represent important post-Born Oppenheimer benchmarks.

Looking forward, because $\bm \Gamma_{\mu \nu}$ is a one-electron operator, diagonalizing  $H_{PS}(\bm X,\bm P)$  requires the same computational cost as solving a standard electronic Hamiltonian $H_{el}(\bm X)$ as far as the  electronic structure is concerned. 
 Thus, the phase-space electronic approach proposed here offers a physically meaningful as well as computationally practical framework 
for recovering coupled nonadiabatic nuclear-electronic-spin dynamics, going beyond electrostatic studies of non-Born-Oppenheimer dynamic and establishing a crucial link between the chemical dynamics and spintronic problems.
 

\section{Acknowledgements}
We thank 
Al Viggiano, Nick Schuman, and Shaun Ard for focusing our attention on the problem of angular momentum conservation
and we thank Clàudia Climent, and Vishikh Athavale for helpful conversations
in developing the relevant matrix elements and equations of motion. 
This work is supported by the National Science Foundation under Grant No. CHE-2102402.

\appendix
\section{Rotations of Atomic Orbital Matrix Elements}
\label{app:ao_rot}

\subsection{Overview}
In this appendix, we will derive the fundamental (intuitive) rules that govern how the one and two electron operators transform under rotation, as well as how the linear and angular momentum transform. Let ${\bm R}$ be a rotational operator in  Cartesian $xyz$ space that rotates around the axis  $\bm{\delta}/\left|\bm{\delta}\right|$ by an amount $\left|\bm{\delta}\right|$: 
$\bm{R} = \mathrm{exp}(-\frac{i}{\hbar}\sum_{\alpha}\bm{L}^{\alpha}\delta_{\alpha})$. Let $\ket{{\nu}_{B}}$ be a basis function on atomic center B and let $\ket{\bar{\nu}_{B}}$ be a rotation of that basis function around center B ,$\ket{\bar{\nu}_{B}} = \mathrm{exp}(-\frac{i}{\hbar}\sum_{\alpha}\hat{L}^{B\alpha}_{e}\delta_{\alpha})\ket{\nu_{B}}$, where
\begin{eqnarray}
\label{eq:LA_hat}
  \hat{L}^{B\alpha}_{e}= -i\hbar \sum_{\beta\gamma}\epsilon_{\alpha\beta\gamma}(r_{\beta}-R_{B\beta})\frac{\partial}{\partial  r^{\gamma}} =  \hat{L}_{e}^{\alpha} -  \sum_{\beta\gamma}\epsilon_{\alpha\beta\gamma}X_{B\beta}\hat{p}^{\gamma}_{e}
\end{eqnarray}  
We will prove:
\begin{eqnarray}
\label{eq:h_rot}
    h_{\bar{\mu}\bar{\nu}}(\bm{R} \bm X_{0}) &=& h_{\mu\nu}(\bm X_{0})
\\
    \label{eq:p_rot}
    p_{\bar{\mu}\bar{\nu}}(\bm{R} \bm X_{0}) &=& \bm{R} p_{\mu\nu}(\bm X_{0})
    \\
    \label{eq:J_rot}
    J_{\bar{\mu}\bar{\nu}}(\bm{R} \bm X_{0}) &=& \bm{R} J_{\mu\nu}(\bm X_{0})
    \\
    \label{eq:pi_rot}
        \pi_{\bar{\mu}\bar{\nu}\bar{\lambda}\bar{\sigma}}(\bm{R} \bm X_{0}) &=&\pi_{\mu\nu\lambda\sigma} (\bm X_{0})
\end{eqnarray}

These relationships are broadly important both for the results in this manuscript as well as those in Ref. \citenum{Tian2023:ERF}. We will also prove  below the equivalence of Eq.\ref{eq:Gamma_uv4} and Eq.\ref{eq:Gamma_uv4prime}.

\subsection{Overview of One-Electron Operator Atomic Orbital Matrix Elements}
Let us write the matrix elements for general one-electron operator $\hat{\mathcal{O}}^{1}$
in atom-centered basis as $\mathcal{O}^{1}_{\mu_{A}\nu_{B}}(\bm X_{0})$, where  basis function $\mu$ is centered on atom A and $\nu$ is centered on atom B. We imagine performing two rotations: $(i)$ first we rotate the whole molecule $\bm{R} \bm X_{0}$ and $(ii)$ second we rotate the basis functions around their atomic centers. 
We wish to evaluate the first order change of the matrix element of the general one-electron operator ${\mathcal{O}}^{1}_{\bar{\mu}_{A}\bar{\nu}_{B}}(\bm{R} \bm X_{0}) $:
\begin{align}
    \zeta(\bm\delta)  = {\mathcal{O}}^{1}_{\bar{\mu}_{A}\bar{\nu}_{B}}(\bm{R} \bm X_{0})  - \mathcal{O}^{1}_{\mu_{A}\nu_{B}}(\bm X_{0}) - O(\bm\delta^2) 
\end{align}
We will compute the contributions of the two rotations  above to $ \zeta(\bm\delta)$ separately. 
\begin{enumerate}
    \item If we ignore the change in orbitals, the first order change in the matrix elements  due to the molecular rotation $\bm{R} \approx 1 - \frac{i}{\hbar} \sum_{\alpha}\bm{L}^{\alpha}\delta_{\alpha} $ is: 
    \begin{eqnarray}
    \label{eq:1e_rot_1}
 {\mathcal{O}}^{1}_{\mu_{A}\nu_{B}}(\bm{R} \bm X_{0})  - \mathcal{O}^{1}_{\mu_{A}\nu_{B}}(\bm X_{0})  
     \approx
    \frac{-i}{\hbar}\sum_{C\alpha\beta\gamma}\delta_{\alpha}\frac{\partial {\mathcal{O}}^{1}_{\mu_{A}\nu_{B}}}{\partial X_{C\alpha}}(L^{\alpha}_{\beta\gamma}X_{C\gamma}) = \sum_{C\alpha\beta\gamma} \delta_{\alpha}\epsilon_{\beta\gamma\alpha} X_{C\gamma} \frac{\partial {\mathcal{O}}^{1}_{\mu_{A}\nu_{B}}}{\partial X_{C\alpha}}
\end{eqnarray}
Here we have used the relationship $L^{\alpha}_{\beta\gamma} = i\hbar \epsilon_{\alpha\beta\gamma}$. We can further expand the nuclear derivatives of the one-electron matrix elements into three components:
\begin{eqnarray}
\label{eq:1e_R}
    \frac{\partial {\mathcal{O}}^{1}_{\mu_{A}\nu_{B}}}{\partial X_{C\gamma}} = \bra{\mu_{A}} \frac{\partial \hat{\mathcal{O}}^{1}}{\partial X_{C\gamma}}\ket{\nu_{B}} + \bra{\frac{\partial }{\partial X_{A\gamma}}\mu_{A}} \hat{\mathcal{O}}^{1}\ket{\nu_{B}}\delta_{AC} + \bra{\mu_{A}} \hat{\mathcal{O}}^{1}\ket{\frac{\partial }{\partial X_{B\gamma}}\nu_{B}}\delta_{BC}
\end{eqnarray}
    \item Next, we compute the first order change due to rotating the atomic basis functions around their centers. Considering the first order change in $\delta_{\alpha}$ for $\ket{\bar{\nu}_{B}}$ for example, we find:
    \begin{eqnarray}
        \ket{\bar{\nu}_{B}} \approx \Big(1-\frac{i}{\hbar} \sum_{\alpha}\hat{L}^{B\alpha}_{e} \delta_{\alpha}   \Big)\ket{{\nu}_{B}} 
    \end{eqnarray}
    If we plug in  Eq. \ref{eq:LA_hat}, we can evaluate the contribution to the first order change:
    \begin{eqnarray}
    {\mathcal{O}}^{1}_{{\mu}_{A}\bar{\nu}_{B}}(\bm{R} \bm X_{0}) 
        -{\mathcal{O}}^{1}_{\mu_{A}\nu_{B}}(\bm{R} \bm X_{0}) \approx
        \frac{-i}{\hbar}\sum_{\alpha}\delta_{\alpha}\Bigg( \bra{\mu_{A}} \hat{\mathcal{O}}^{1}\ket{\hat{L}^{\alpha}_{e} \nu_{B}} - \sum_{\beta\gamma}\epsilon_{\alpha\beta\gamma}X_{B\beta} \bra{ \mu_{A}}\hat{\mathcal{O}}^{1}\ket{\hat{p}^{\gamma}_{e} \nu_{B}} \Bigg)
    \end{eqnarray}
    Now if we also add in the first order change in $\delta_{\alpha}$ for $\bra{\bar{\mu}_{A}}$, we recover
    \begin{eqnarray}
    \label{eq:1e_rot_2}
    & &{\mathcal{O}}^{1}_{\bar{\mu}_{A}\bar{\nu}_{B}}(\bm{R} \bm X_{0}) 
       -{\mathcal{O}}^{1}_{\mu_{A}\nu_{B}}(\bm{R} \bm X_{0})   \\
       \nonumber
       & \approx &  \frac{-i}{\hbar}\sum_{\alpha}\delta_{\alpha}\Bigg[
        \bra{\mu_{A}} [\hat{\mathcal{O}}^{1},\hat{L}^{\alpha}_{e} ]\ket{\nu_{B}} + \sum_{\beta\gamma}\epsilon_{\alpha\beta\gamma}\Big(X_{A\beta}\bra{ \hat{p}^{\gamma}_{e} \mu_{A}}\hat{\mathcal{O}}^{1}\ket{\nu_{B}} -X_{B\beta} \bra{ \mu_{A}}\hat{\mathcal{O}}^{1}\ket{\hat{p}^{\gamma}_{e} \nu_{B}} \Big)\Bigg]
    \end{eqnarray}
    Finally, if we use the relationship $(\hat{p}^{\gamma}_{e}  -i\hbar\nabla^{\gamma}_{n}) \ket{ \nu_{B}}=0$, the second term in Eq. \ref{eq:1e_rot_2} becomes 
    \begin{eqnarray}\label{eq:1e_rot_2_1}
        &&\frac{-i}{\hbar}\sum_{\alpha\beta\gamma}\delta_{\alpha}\epsilon_{\alpha\beta\gamma}\Big(X_{A\beta}\bra{ \hat{p}^{\gamma}_{e} \mu_{A}}\hat{\mathcal{O}}^{1}\ket{\nu_{B}} -X_{B\beta} \bra{ \mu_{A}}\hat{\mathcal{O}}^{1}\ket{\hat{p}^{\gamma}_{e} \nu_{B}} \Big) \nonumber \\
&=&-\sum_{\alpha\beta\gamma}\delta_{\alpha}\epsilon_{\alpha\beta\gamma}\Bigg(X_{A\beta}\bra{\frac{\partial \mu_{A}}{\partial X_{A\gamma}}} \hat{\mathcal{O}}^{1} \ket{\nu_{B}} + X_{B\beta}\bra{\mu_{A}} \hat{\mathcal{O}}^{1} \ket{\frac{\partial \nu_{B}}{\partial X_{B\gamma}}}\Bigg)
    \end{eqnarray}
\end{enumerate}
Finally,  adding these two contributions together and noting that the last two terms in Eq. \ref{eq:1e_R} cancel with the terms in Eq. \ref{eq:1e_rot_2_1},  we find:
\begin{eqnarray}
\label{eq:1e_Rbar}
\zeta(\bm\delta)=
\sum_{C\alpha\beta\gamma}\delta_{\alpha} \epsilon_{\alpha\beta\gamma} X_{C\beta} \bra{\mu_{A}} \frac{\partial \hat{\mathcal{O}}^{1}}{\partial X_{C\gamma}}\ket{\nu_{B}} -\frac{i}{\hbar}\delta_{\alpha}\sum_{\alpha}
        \bra{\mu_{A}} [\hat{\mathcal{O}}^{1},\hat{L}^{\alpha}_{e} ]\ket{\nu_{B}} 
\end{eqnarray}
Eq. \ref{eq:1e_Rbar} is general and allows us to evaluate the first order changes under rotation for different one-electron matrix elements.

\subsubsection{One-Electron Core Hamiltonian Matrix Elements (Proving Eq.\ref{eq:h_rot})}
    Due to the isotropy of the space, the following commutator is zero:
    \begin{eqnarray}
    \label{eq:h_commu}
        \left[\hat{h}, \hat{L}^{\alpha}_{e} + \hat{L}^{\alpha}_{n}\right] = 0
    \end{eqnarray}
    Plugging this commutation relation into Eq. \ref{eq:1e_Rbar}, the second term on the RHS becomes 
    \begin{eqnarray}\label{eq:h_rot_commu}
        \frac{-i}{\hbar}\sum_{\alpha}\delta_{\alpha}\bra{\mu_{A}} [\hat{h},\hat{L}^{\alpha}_{e} ]\ket{\nu_{B}} = \frac{i}{\hbar}\sum_{\alpha}\delta_{\alpha}\bra{\mu_{A}} [\hat{h},\hat{L}^{\alpha}_{n} ]\ket{\nu_{B}} =-\sum_{C\alpha\beta\gamma} \delta_{\alpha} \epsilon_{\alpha\beta\gamma} X_{C\beta} \bra{\mu_{A}} \frac{\partial \hat{h}}{\partial X_{C\gamma}}\ket{\nu_{B}}
    \end{eqnarray}
        Plugging Eq. \ref{eq:h_rot_commu} back to Eq. \ref{eq:1e_Rbar}, we see that two terms cancel with each other, thus proving Eq. \ref{eq:h_rot}.

\subsubsection{Electronic Linear Momentum Matrix Elements (Proving Eq.\ref{eq:p_rot})}
We begin by evaluating the commutator,
\begin{eqnarray}
    \left[\hat{p}^{\eta}_{e}, \hat{L}^{\alpha}_{e}\right] = i\hbar\sum_{\gamma}\epsilon_{\eta\alpha\gamma}\hat{p}^{\gamma}_{e}
\end{eqnarray}
Plugging this commutation relation in Eq. \ref{eq:1e_Rbar}, we find  
\begin{align}
\zeta(\bm\delta)= 
\sum_{\alpha\gamma}\delta_{\alpha}\epsilon_{\eta\alpha\gamma}
        \bra{\mu_{A}} \hat{p}^{\gamma}_{e}\ket{\nu_{B}}
\end{align}
which is the first order change corresponding to the RHS of Eq. \ref{eq:p_rot}.

\subsubsection{$J_{\mu_{A}\nu_{B}}$ Matrix Elements  (Proving Eq.\ref{eq:J_rot})}
We first review the definition of the $J^{\eta}_{\mu_{A}\nu_{B}}$ matrix elements,
\begin{eqnarray}
    J^{\eta}_{\mu_{A}\nu_{B}} = -\frac{i}{2\hbar} \bra{\mu_{A}}\left(\hat{{L}}^{A\eta}_{e}+\hat{{L}}^{B\eta}_{e}\right)\ket{\nu_{B}}
\end{eqnarray}
Now we evaluate the commutation relation, 
\begin{eqnarray}
    [\hat{{l}}^{A\eta},\hat{L}^{\alpha}_{e}] = [\hat{L}^{\eta}_{e} - \sum_{\beta\gamma}\epsilon_{\eta\beta\gamma}X_{A\beta}\hat{p}^{\gamma}_{e},\hat{L}^{\alpha}_{e}] = i\hbar\sum_{\gamma}\epsilon_{\eta\alpha\gamma}\hat{L}^{\gamma}_{e} - i\hbar\sum_{\beta\gamma\kappa}\epsilon_{\eta\beta\gamma}X_{A\beta}\epsilon_{\gamma\alpha\kappa}\hat{p}^{\kappa}_{e}
\end{eqnarray}
Plugging this commutation relation in Eq. \ref{eq:1e_Rbar}, we find
\begin{eqnarray}
\label{eq:L_rot_1}
   \zeta(\bm\delta) &=&  -\frac{i}{2\hbar}\sum_{\alpha\beta\gamma}\delta_{\alpha} \epsilon_{\alpha\beta\kappa} X_{A\beta}\epsilon_{
    \eta\kappa\gamma} \bra{\mu_{A}}\hat{p}^{\gamma}_{e} \ket{\nu_{B}} -\frac{i}{2\hbar}\sum_{\alpha\gamma}\delta_{\alpha}\epsilon_{\eta\alpha\gamma}
        \bra{\mu_{A}} \hat{L}^{\gamma}_{e} \ket{\nu_{B}} \nonumber\\
        &&+ \frac{i}{2\hbar}\sum_{\alpha\beta\gamma\kappa}\delta_{\alpha}\epsilon_{\eta\beta\gamma}X_{A\beta}\epsilon_{\gamma\alpha\kappa}\bra{\mu_{A}}\hat{p}^{\kappa}_{e}\ket{\nu_{B}}
\end{eqnarray}
This is the first order change for the term $-\frac{i}{2\hbar}\bm{R} \bra{\mu_{A}}  \hat{{\bm L}}^{A}\ket{\nu_{B}}$. One can obtain the equivalent result  for the $-\frac{i}{2\hbar}\bm{R} \bra{\mu_{A}}  \hat{{\bm L}}^{B}\ket{\nu_{B}}$ term. Hence we have proven Eq. \ref{eq:J_rot}.

\subsection{Two-Electron Operator Atomic Orbital Matrix Elements (Proving Eq.\ref{eq:pi_rot})}
The evaluation of the first order changes in  the two-electron operator matrix elements upon rotation is very similar to the approach above. Given a general two-electron operator $\hat{\mathcal{O}}^{2}$, the first order changes $\zeta(\bm \delta)$ to  the matrix elements $c$ upon rotation can be evaluated as follows:
\begin{itemize}
    \item The first order change due to the molecular rotation $\bm{R} \approx 1 - \frac{i}{\hbar} \sum_{\alpha}\bm{L}^{\alpha}\delta_{\alpha} $ (similar to Eq. \ref{eq:1e_rot_1}) is
    \begin{eqnarray}
{\mathcal{O}}^{2}_{\mu_{A}\nu_{B}\lambda_{C}\sigma_{D}}(\bm{R} \bm X_{0})  - \mathcal{O}^{2}_{\mu_{A}\nu_{B}\lambda_{C}\sigma_{D}}(\bm X_{0})  
    & \approx&   \label{eq:2e_rot_1}
    \frac{-i}{\hbar}\sum_{Q\alpha\beta\gamma}\delta_{\alpha}\frac{\partial {\mathcal{O}}^{2}_{\mu_{A}\nu_{B}\lambda_{C}\sigma_{D}}}{\partial X_{C\alpha}}(L^{\alpha}_{\beta\gamma}X_{Q\gamma})\\
    &=& \sum_{Q\alpha\beta\gamma} \delta_{\alpha}\epsilon_{\beta\gamma\alpha} X_{Q\gamma} \frac{\partial {\mathcal{O}}^{2}_{\mu_{A}\nu_{B}\lambda_{C}\sigma_{D}}}{\partial X_{Q\alpha}}
\end{eqnarray}
 We can further expand the nuclear derivatives of the two-electron matrix elements into five components:
\begin{eqnarray}
\label{eq:2e_R}
    \frac{\partial {\mathcal{O}}^{2}_{\mu_{A}\nu_{B}\lambda_{C}\sigma_{D}}}{\partial X_{Q\gamma}} &=& \bra{\mu_{A}\nu_{B} }\frac{\partial \hat{\mathcal{O}}^{2}}{\partial X_{Q\gamma}}\ket{\lambda_{C}\sigma_{D}} + \bra{\Big(\frac{\partial }{\partial X_{A\gamma}}\mu_{A}\Big)\lambda_{C}\sigma_{D}} \hat{\mathcal{O}}^{2}\ket{\nu_{B}}\delta_{AG}   \nonumber \\
   &&+ \bra{\mu_{A}\frac{\partial }{\partial X_{B\gamma}}\nu_{B}} \hat{\mathcal{O}}^{2}\ket{\lambda_{C}\sigma_{D}}\delta_{BG} + \bra{\mu_{A}\nu_{B} }\hat{\mathcal{O}}^{2}\ket{\Big(\frac{\partial }{\partial X_{C\gamma}}\lambda_{C}\Big)\sigma_{D}}\delta_{CG} \nonumber\\
   &&+ \bra{\mu_{A}\nu_{B} }\hat{\mathcal{O}}^{2}\ket{\lambda_{C}\frac{\partial }{\partial X_{D\gamma}}\sigma_{D}}\delta_{DG}
\end{eqnarray}
\item Similar to Eq. \ref{eq:1e_rot_2}, the first order change due to the basis functions rotating around their centers can be written as:
\begin{eqnarray}
    \label{eq:2e_bar_1}
    & & 
\hat{\mathcal{O}}^{2}_{\bar{\mu}_{A}\bar{\nu}_{B}\bar{\lambda}_{C}\bar{\sigma}_{D}} - \mathcal{O}^{2}_{\mu_{A}\nu_{B}\lambda_{C}\sigma_{D}}(\bm X_{0})  
     \approx    
        \frac{-i}{\hbar}\sum_{\alpha n}\delta_{\alpha}
        \bra{\mu_{A}\nu_{B}} [\hat{\mathcal{O}}^{2},\hat{L}^{\alpha}_{e,n} ]\ket{\lambda_{C}\sigma_{D}} 
\\
   & & 
        -\frac{i}{\hbar}\sum_{\alpha\beta\gamma n}\delta_{\alpha}\epsilon_{\alpha\beta\gamma}\Big[X_{A\beta}\bra{ \hat{p}^{\gamma}_{e,n} \Big(\mu_{A}\nu_{B}\Big)}\hat{\mathcal{O}}^{2}\ket{\lambda_{C}\sigma_{D}} -X_{B\beta} \bra{ \mu_{A}\nu_{B}}\hat{\mathcal{O}}^{2}\ket{\hat{p}^{\gamma}_{e,n} \Big(\lambda_{C}\sigma_{D}}\Big) \Big]
        \nonumber 
    \end{eqnarray}
\end{itemize}
Here the letter $n$ represents the electron number. For a two-electron operator, $n$ can be either electron 1 or 2. When $n$ refers to electron 1, $\hat{L}^{\alpha}_{e}$ acts on $\mu_{A}$ or $\lambda_{C}$. 
We note that the long second term on the RHS of  Eq. \ref{eq:2e_bar_1}  cancels with the terms that involve nuclear derivatives of the basis functions in Eq. \ref{eq:2e_R}. Hence, the first order change in the  two-electron operator upon rotation is
\begin{eqnarray}
\label{eq:2e_Rbar}
   \zeta(\bm \delta)= \sum_{Q\alpha\beta\gamma} \delta_{\alpha}\epsilon_{\beta\gamma\alpha} X_{Q\gamma} \bra{\mu_{A}\nu_{B} }\frac{\partial \hat{\mathcal{O}}^{2}}{\partial X_{Q\alpha}}\ket{\lambda_{C}\sigma_{D}}-  \frac{i}{\hbar}\sum_{\alpha n}\delta_{\alpha}
        \bra{\mu_{A}\nu_{B}} [\hat{\mathcal{O}}^{2},\hat{L}^{\alpha}_{e,n} ]\ket{\lambda_{C}\sigma_{D}} 
\end{eqnarray}
\subsubsection{$\pi_{\mu_{A}\nu_{B}\lambda_{C}\sigma_{D}}$ Matrix Elements}
From the form of the 2-electron interaction operator $\frac{1}{|\hat{\bm{r}}_{1}-\hat{\bm{r}}_{2}|}$, we can easily deduce that the first term in Eq. \ref{eq:2e_Rbar} is zero. For the second term in Eq. \ref{eq:2e_Rbar}, note that
\begin{align}
    \left[\hat{\bm{r}}_1 \times \hat{\bm{p}}_1 + \hat{\bm{r}}_2 \times  \hat{\bm{p}}_2, \frac{1}{\left| \hat{\bm{r}}_1 - \hat{\bm{r}}_2\right|} \right]  = 0 
\end{align}
Hence the second term in Eq. \ref{eq:2e_Rbar} is also zero. Hence we have proven Eq. \ref{eq:pi_rot}.

\section{Cartesian Coordinates in Bohr}
\label{app:coord}
\subsection{Equilibrium Geometries}
The equilibrium geometries for the molecules investigated in Tables \ref{tab:translation}-\ref{tab:rotation_gamma} were optimized with restricted Hartree-Fock and a cc-pVTZ basis set. 
\begin{itemize}
    \item \ce{H_{2}} 
\begin{table}[H]  
\begin{tabular}{c c c c} 
\hline
    H  & -0.693827279423610        &0.000000000000000   &  0.000000000000000 \\
    H  & 0.693827279423610        &0.000000000000000   &  0.000000000000000\\
    \hline
\end{tabular} 
\end{table}   
    \item \ce{LiH} 
\begin{table}[H]  
\begin{tabular}{c c c c} 
\hline
    Li  & -0.381507444748606       &0.000000000000000   &  0.000000000000000 \\
    H  & 2.655860841963647       &0.000000000000000   &  0.000000000000000\\
    \hline
\end{tabular} 
\end{table}   
\item  \ce{HCN}
\begin{table}[H]  
\begin{tabular}{c c c c} 
\hline
    H  & -3.023887918951979       &0.000000000000000   &  0.000000000000000 \\
    C  & -1.027314918951979      &0.000000000000000   &  0.000000000000000\\
    N   & 1.097998081048021     &0.000000000000000   &  0.000000000000000\\
    \hline
\end{tabular} 
\end{table}  
\item  \ce{H_{2}O} 
\begin{table}[H]  
\begin{tabular}{c c c c} 
\hline
    H  & 1.777459680682990       &0.000000000000000   &  0.000000000000000 \\
    O  & 0.000000000000000     &0.000000000000000   &  0.000000000000000\\
    H   & -0.489811956048840     &-1.708638980055550   &  0.000000000000000\\
    \hline
\end{tabular} 
\end{table}  
\item \ce{C_{4}H_{2}} 
\begin{table}[H]  
\begin{tabular}{c c c c} 
\hline
    H  & -5.534109801000000       &0.000000000000000   &  0.000000000000000 \\
    C  & -3.542659862000000     &0.000000000000000   &  0.000000000000000\\
    C   & -1.308625798000000     &0.000000000000000  &  0.000000000000000\\
    C   & 1.308625798000000     &0.000000000000000  &  0.000000000000000\\
    C  & 3.542659862000000     &0.000000000000000   &  0.000000000000000\\
    H  & 5.534109801000000       &0.000000000000000   &  0.000000000000000 \\
    \hline
\end{tabular} 
\end{table}  
\end{itemize}
\subsection{ \ce{LiH} Dimer}
\begin{table}[H]  
\begin{tabular}{c c c c} 
\hline
    Li  & -0.381507444748606       &0.000000000000000   &  0.000000000000000 \\
    H  & 2.655860841963647       &0.000000000000000   &  0.000000000000000\\
    Li & 35.00000000000000 & -0.381507444748606       &0.000000000000000   \\
    H & 35.00000000000000  & 2.655860841963647       &0.000000000000000   \\
    \hline
\end{tabular} 
\end{table}   

\section{ETF and ERF Contributions}
\label{app:gamma_contributions}
\subsection{Linear Momentum When Stretching an H Atom of a Water Molecule }
For all the molecules studied in Table \ref{tab:vibration}, the contributions of $\bm\Gamma''$ are non-zero only for the water molecule. In Table \ref{tab:h2o_gamma}, we show the linear momentum as calculated with either $\bm\Gamma'$ or $\bm\Gamma''$ exclusively. Here, the water molecule is placed in the xy plane with one of the O-H bonds aligned along the x-axis. Moving the H atom away from the O atom along the x-axis puts the majority of the linear momentum in the x direction (<$\hat{p}_e^{x}$>) and a small degree in the y direction (<$\hat{p}_e^{y}$>). Comparing the contributions from the two components of the $\bm \Gamma$ coupling ($\bGamma'$ vs $\bGamma''$), we notice that the contribution from $\bm\Gamma''$ to the linear momentum is much smaller (by an order of magnitude) as compared to the contribution from $\bm\Gamma'$. 

\begin{table}[H]
\centering
\caption{The electronic linear momentum ($\hbar/a_{0}$) calculated with the phase-space Hamiltonian approach with two choices of $\bm \Gamma$: the electron translation factor ($\bm \Gamma'$) and the electron rotation factor ($\bm  \Gamma''$). The velocity used here corresponds to stretching one H atom along the $x$ axis of the water molecule. }
\label{tab:h2o_gamma}
\renewcommand{\arraystretch}{2}
\begin{tabular}{ccccccccc}
\hline
 & & STO-3G &DZ & aDZ & TZ & aTZ & QZ & aQZ  \\ \hline
 <$\hat{p}_e^{x}$> & $\bm\Gamma'$ & 3.57e-4 & 7.11e-4 & 6.67e-4 & 2.75e-4 &5.13e-4&2.03e-4&7.74e-4 \\
 & $\bm\Gamma''$& 1.21e-5&4.18e-5&-9.90e-6&6.20e-5&-1.52e-6&7.40e-5&7.38e-6\\
<$\hat{p}_e^{y}$>  & $\bm\Gamma'$  & 5.82e-6&-6.86e-5&-4.20e-5&-1.08e-4&-1.21e-5&-1.96e-4&-7.21e-5\\
& $\bm\Gamma''$  & 9.15e-6&3.15e-5&-7.46e-6&4.67e-5&-1.17e-6&-5.57e-5&4.93e-6\\
 \hline
\end{tabular} \\
\end{table}

\subsection{Angular Momentum of Rigid Rotations of Linear Molecules}
In this section, we provide data regarding the angular momentum as calculated using either  $\bm \Gamma'$ and $\bm \Gamma''$ exclusively. Depending on the molecules, for the equilibrium geometries, the angular momentum calculated with only $\bm \Gamma''$ can be the same magnitude or one magnitude smaller than what is found when using $\bm\Gamma'$ alone. At the geometry for which the diatomic bond lengths are stretched to 8 Bohr, the contribution from $\bm\Gamma'$ clearly dominates the contribution from $\bm\Gamma''$. 
\begin{table}[H]
\centering
\caption{The electronic angular momentum <$\hat{L}_e^{z}$> ($\hbar$) calculated with the phase-space Hamiltonian approach with the electron translation factor ($\bm \Gamma'$) or the electron rotation factor ($\bm  \Gamma''$) only. The velocity used here corresponds to a rigid rotation around the $z$ axis of the linear molecules aligned at the $x$ axis. }
\label{tab:rotation_gamma}
\renewcommand{\arraystretch}{2}
\begin{tabular}{ccccccccc}
\hline
 & & STO-3G &DZ & aDZ & TZ & aTZ & QZ & aQZ  \\ \hline
\ce{H_{2}} & $\bm\Gamma'$ & 0.00& 1.37e-4& 3.26e-4 & 2.40e-4 &2.47e-4&2.88e-4&2.07e-4 \\
 & $\bm\Gamma''$& 0.00&-8.40e-5&-2.51e-5&-1.68e-4&-1.75e-4&-2.15e-4&-1.34e-4\\
Stretched \ce{H_{2}}  & $\bm\Gamma'$  & 0.00&6.93e-3&2.33e-2&1.35e-2&2.46e-2&1.72e-2&2.50e-2\\
& $\bm\Gamma''$  & 0.00&-1.70e-5&-6.68e-4&-9.67e-5&-9.85e-4&-2.02e-4&-1.15e-3\\
\ce{LiH}  & $\bm\Gamma'$  & 9.11e-4&3.50e-3&6.22e-3&4.93e-3&7.55e-3&5.77e-3&6.85e-3\\
& $\bm\Gamma''$  &4.83e-3&5.89e-3&4.47e-3&5.73e-3&3.64e-3&5.34e-3&4.51e-3\\
\ce{HCN}  & $\bm\Gamma'$  & 3.33e-4&1.96e-3&2.41e-3&2.71e-3&2.66e-3&2.98e-3&2.64e-3\\
& $\bm\Gamma''$  & 2.14e-3&1.53e-3&1.28e-3&1.27e-3&1.40e-3&1.27e-3&1.63e-3\\
\ce{C_{4}H_{2}}  & $\bm\Gamma'$  & 5.78e-4&8.17e-3&1.18e-2&9.74e-3&9.76e-3&1.07e-2&1.07e-2\\
& $\bm\Gamma''$  & 9.60e-4&-6.76e-4&-2.97e-3&-2.90e-5&5.58e-4&2.40e-4&3.57e-4\\
 \hline
\end{tabular} \\
\end{table}

\bibliography{Ref}

\begin{mcitethebibliography}{103}
\bibitem[joe()]{joefootnote}
As a side note, we mention that the ETFs described in the present paper correspond exactly to velocity-gauge factors in the NVP literature, but there is no analogue to ERFs in the latter.\relax
\mciteBstWouldAddEndPunctfalse
\mciteSetBstMidEndSepPunct{\mcitedefaultmidpunct}
{}{\mcitedefaultseppunct}\relax
\EndOfBibitem
\end{mcitethebibliography}

\end{document}